\documentclass[twocolumn]{aastex61}
\usepackage{tikz}
\usetikzlibrary{shapes.geometric, arrows}
\usetikzlibrary{calc}
\usepackage[]{algorithm}
\usepackage{amsmath}
\usepackage{graphicx}
\usepackage{subfigure}
\usepackage{threeparttable}
\usepackage{float}
\usepackage{todonotes}

\received{December 12 2017}
\revised{June 8 2018}
\accepted{August 5 2018}

\submitjournal{ApJ}

\shorttitle{Hybrid Foreground Removal Technique}
\shortauthors{Kerrigan et al.}

\begin{document}

\title{Improved 21cm Epoch of Reionization Power Spectrum Measurements with a Hybrid Foreground Subtraction and Avoidance Technique}

\correspondingauthor{Joshua R. Kerrigan}
\email{joshua\_kerrigan@brown.edu}

\author{Joshua R. Kerrigan}
\affil{Department of Physics, Brown University, Providence RI}

\author{Jonathan C. Pober}
\affil{Department of Physics, Brown University, Providence RI}

\author{Zaki S. Ali}
\affil{Department of Astronomy, University of California, Berkeley CA}

\author{Carina Cheng}
\affil{Department of Astronomy, University of California, Berkeley CA}

\author{Adam P. Beardsley}
\affil{School of Earth and Space Exploration, Arizona State University, Tempe AZ}

\author{Aaron R. Parsons}
\affil{Department of Astronomy, University of California, Berkeley CA}
\affil{Radio Astronomy Lab., University of California, Berkeley CA}

\author{James E. Aguirre}
\affil{Department of Physics and Astronomy, University of Pennsylvania, Philadelphia PA}

\author{Nichole Barry}
\affil{Department of Physics, University of Washington, Seattle WA}

\author{Richard F. Bradley}
\affil{Department of Electrical and Computer Engineering, University of Virginia, Charlottesville VA}
\affil{National Radio Astronomy Observatory, Charlottesville VA}
\affil{Department of Astronomy, University of Virginia, Charlottesville VA}

\author{Gianni Bernardi}
\affil{INAF - Istituto di Radioastronomia, Italy}
\affil{Square Kilometer Array, South Africa, Cape Town South Africa}
\affil{Department of Physics and Electronics, Rhodes University, South Africa}

\author{Chris L. Carilli}
\affil{National Radio Astronomy Observatory, Socorro NM}
\affil{Astrophysics Group, Cavendish Lab., Cambridge UK}

\author{David R. DeBoer}
\affil{Radio Astronomy Lab., University of California, Berkeley CA}

\author{Joshua S. Dillon} \altaffiliation{NSF Astronomy and Astrophysics Postdoctoral Fellow}
\affil{Department of Astronomy, University of California, Berkeley CA}
\affil{Radio Astronomy Lab., University of California, Berkeley CA}
\affil{Berkeley Center for Cosmological Physics, Berkeley CA}

\author{Daniel C. Jacobs}
\affil{School of Earth and Space Exploration, Arizona State University, Tempe AZ}

\author{Saul A. Kohn}
\affil{Department of Physics and Astronomy, University of Pennsylvania, Philadelphia PA}

\author{Matthew Kolopanis}
\affil{School of Earth and Space Exploration, Arizona State University, Tempe AZ}
\affil{Department of Physics, Arizona State University, Tempe AZ}

\author{Adam Lanman}
\affil{Department of Physics, Brown University, Providence RI}

\author{Wenyang Li}
\affil{Department of Physics, Brown University, Providence RI}

\author{Adrian Liu} \altaffiliation{Hubble Fellow}
\affil{Department of Astronomy, University of California, Berkeley CA}
\affil{Berkeley Center for Cosmological Physics, Berkeley CA}

\author{Ian Sullivan}
\affil{Department of Physics, University of Washington,  Seattle WA}
 
\begin{abstract}
Observations of the 21cm Epoch of Reionization (EoR) signal are dominated by Galactic and extragalactic foregrounds. The need for foreground removal has led to the development of two main techniques, often referred to as ``foreground avoidance" and ``foreground subtraction." Avoidance is associated with filtering foregrounds in Fourier space, while subtraction uses an explicit foreground model that is removed. Using 1088 hours of data from the 64-element PAPER array, we demonstrate that subtraction of a foreground model prior to delay-space foreground filtering results in a modest but measurable improvement of the performance of the filter. This proof-of-concept result shows that improvement stems from the reduced dynamic range requirements needed for the foreground filter: subtraction of a foreground model reduces the total foreground power, so for a fixed dynamic range, the filter can push towards fainter limits. We also find that the choice of window function used in the foreground filter can have an appreciable affect on the performance near the edges of the observing band. We demonstrate these effects using a smaller 3 hour sampling of data from the MWA, and find that the hybrid filtering and subtraction removal approach provides similar improvements across the band as seen in the case with PAPER-64.
\end{abstract}


\section{Introduction} \label{sec:intro}
The 21cm Epoch of Reionization (EoR) signal is faint emission produced from neutral hydrogen in the period directly following the formation of the first stars and galaxies. We can attempt to observe this redshift- evolving signal by using radio interferometer arrays with a wide instantaneous bandwidth. Unfortunately, we are limited by what we can observe due in part to Galactic and extragalactic foregrounds concealing the 21cm signal in our observations. First generation 21cm interferometers such as the Murchison Widefield Array (MWA) \citep{Tingay2013}, the Donald C. Backer Precision Array for Probing the Epoch of Reionization (PAPER) \citep{Parsons2010}, and the LOw Frequency Array (LOFAR) \citep{vanHaarlem2013} have each developed independent techniques for addressing these foregrounds in their observations. MWA and LOFAR have developed algorithms to fit and subtract accurate foreground models to observations, while PAPER uses foreground ``avoidance'' by filtering in Fourier space on individual visibilities per baseline and per time. There is, however, no \emph{a priori} reason that these techniques can only be applied to their respective telescopes.

Practically speaking, there are potential limitations. For example, PAPER uses redundant array configurations, and so its ability to accurately model foregrounds is limited by the point spread function (PSF) and $(u,v)$ coverage of the array.
Similarly, the MWA has more complicated structure in its frequency response due to the polyphase filter bank \citep{Offringa2016} and cable reflections \citep{Ewall-Wice2016}. This may require a wider delay-space filter in the MWA case to remove significant foreground power.

In this work, we present a method using both model-based foreground subtraction and delay-space filtering. Applied to both data from PAPER and the MWA, we find improvements in 21cm EoR power spectrum sensitivities over all redshifts. In Section \ref{section:background} we provide motivation for removing foregrounds in our observations and the need for additional foreground suppression. Section \ref{section:wideband} introduces the delay spectrum and filtering technique as applied in past PAPER analyses and describes how it is used in this work. Section \ref{section:fsmethod} describes how the foreground models to be subtracted are processed and applied to PAPER visibilities. In Section \ref{section:data} we provide specifications for the PAPER and MWA data used for this analysis. In Section \ref{section:pipeline} we compare the result of our hybrid technique with both the PAPER power spectrum pipeline as presented in \citet{Ali2015} and a technique using only model-based foreground subtraction. Our technique is additionally applied to MWA observations, the results of which are described in Section \ref{section:mwa}. Finally, in Section \ref{section:conclusions}, we conclude by discussing the implications of using foreground filtering and subtraction for future power spectra analyses.

\section{Background}\label{sec:back}
\label{section:background}

Galactic and extragalactic foregrounds dominate the cosmological signal in 21cm EoR experiments, and must be removed if a detection of the 21cm signal is to be achieved. Several foreground mitigation techniques have been applied independently to arrays such as PAPER, MWA, and LOFAR (see e.g. \citealt{Liu2011,Vedantham2012FG,Parsons2012,Dillon2013FG,Wang2013FG,Parsons2014FG,Liu2014FGA,Liu2014FGB,Dillon2015FG,Pober2016,Trott2016FG}). Some novel examples include a non-parametric GMCA wavelet decomposition \citep{Champman2013} used by LOFAR \citep{Patil2017LOFAR} and an SVD technique that is shown to isolate smooth spectrum foregrounds in GMRT visibilities while making no a priori assumptions about their form \citep{Paciga2013}.

Our point of reference begins with the foreground filtering technique used in \citet{Ali2015} on PAPER-64 observations. This filtering technique uses a wide-band iterative deconvolution filtering approach for removing bright foregrounds, which we describe in more detail in Section \ref{section:wideband}. This filtering however comes at the cost of losing all EoR information at the lowest Fourier modes where foregrounds reside.\footnote{This filtering technique is widely associated with the terminology of ``foreground avoidance.'' The goal is to keep high $k_{\parallel}$ modes (the so-called ``EoR window") of the 21cm power spectrum clear of foreground contamination, at the cost of indiscriminately removing power from low $k_{\parallel}$ modes.  Hence, these low $k$, foreground contaminated modes are ``avoided.''} Foreground subtraction is a method employed by both MWA and LOFAR \citep{Jacobs2016,Beardsley2016,Patil2017LOFAR}, where one simulates visibilities corresponding to an accurate foreground model and subtracts them from the data. Unlike filtering, this method should ideally retain EoR information at all Fourier modes. 


The motivation behind integrating these two techniques into a single power spectrum pipeline is to further reduce band-limited spectral leakage\footnote{Commonly in Digital Signal Processing, spectral leakage is associated with power contaminating neighboring bins in frequency when Fourier transforming from a time series to frequency spectrum. In the work outlined here we refer to spectral leakage as contamination in nearby delay $(\tau)$ bins when transforming from a frequency space.} of foreground power into high cosmological line-of-sight Fourier modes, $k_{\parallel}$, beyond what is capable with a filter alone \citep{Parsons2012,Vendantham2012,Morales2012,Pober2013}. Analogously with a time series of finite data, a windowing function can be applied to reduce spectral leakage. This provides a reasonable way to mitigate any power seeping into neighboring bins in frequency space when taking the Fourier Transform, but it comes at the expense of reduced sensitivity and correlated measurements. Subtracting a foreground model can reduce the dynamic range required in the frequency Fourier transform to
$k_{\parallel}$, mitigating the need for more aggressive windowing functions (e.g. a Blackman-Harris convolved with itself, as in \citealt{Thyagarajan2016}).

\section{Wide-band Delay Filtering for Foreground Avoidance}
\label{section:wideband}
To understand the wide-band delay filtering method employed in PAPER we begin with an explanation of the delay spectrum approach. We use the alternative form of the standard radio interferometer visibility equation
\begin{equation}
\begin{split}
V_{b}(\nu) &= \int dl \ dm \ A(\nu,l,m)I(\nu,l,m)e^{-2\pi i(ul + vm)}
\\&=\int dl \ dm \ A(\nu,l,m)I(\nu,l,m)e^{-2\pi i\nu \tau_{g}}
\label{visibilityeqn}
\end{split}
\end{equation}
where $V_{b}(\nu)$ is the visibility of baseline $b$ at observing frequency $\nu$,  $A(\nu,l,m)$ is the antenna response, $I(\nu,l,m)$ is the specific intensity distribution on the sky, $(l,m)$ are the direction cosines on the sky, and $(u,v)$ are the projected lengths of the baseline in units of wavelength. This form of the interferometer visibility equation is especially useful for demonstrating the delay spectrum approach because we can associate sources on the sky with a geometric delay, $\tau_g$. This geometric delay
\begin{equation}
\tau_{g} = \frac{\vec{b}\cdot \hat{s}}{c} = \frac{1}{c}(b_{x}l + b_{y}m)
\label{delayTau}
\end{equation}
corresponds to the light travel time distance between two antennas for an emission from the direction $\hat{s}$, and directly relates to the baseline lengths $\vec{b} = (b_x,b_y)$ and the visibility domain coordinates with $\vec{u} = \nu \vec{b}/c$.
By then applying a window function $W(\nu)$, and taking the Fourier Transform over frequency of Equation \ref{visibilityeqn},
\small
\begin{equation}
\begin{gathered}
\widetilde{V}_{b}(\tau) =\int^{\infty}_{-\infty} d\nu W(\nu) V_{b}(\nu)e^{-2\pi i\tau\nu} \\
=\int^{\infty}_{-\infty} d\nu W(\nu) \Bigg[\int dl \ dm \ A(\nu,l,m)I(\nu,l,m)e^{-2\pi i\nu \tau_{g}} \Bigg]e^{2\pi i \nu \tau} \\ = 
\int dl \ dm \ \Bigg[\widetilde{W}(\tau) * \widetilde{A}(\tau,l,m)*\widetilde{I}(\tau,l,m)*\delta(\tau-\tau_{g})\Bigg]
\label{delayspectrum}
\end{gathered}
\end{equation}
\normalsize
we have what is called the Delay Transform, as found in \citet{Parsons2009}. This means that point sources on the sky can be mapped from our visibility to a geometric delay in delay space. Flat spectrum  sources are constrained to a maximum geometric delay, $\tau_h$, the horizon limit.\footnote{The intrinsic spectrum of the source expands its footprint in delay space beyond that of a delta function, but smooth spectrum sources (like foregrounds) will still have compact footprints and limited extent beyond the baseline's delay horizon, $\tau_{h}$, which is a direct consequence seen from Equation \ref{delayspectrum}.} This maximum delay is achieved when the direction from the baseline to source on the sky, $\hat{s}$, is parallel to the baseline $\vec{b}$, which means the source is located at the horizon.
When measuring the relatively weak 21cm EoR signal, spectral leakage when taking the Delay Transform is significant. The foreground dominated delays at $|\tau| \leq |\tau_{H}|$ can be approximately $10^{5}$ larger in amplitude than the EoR modes being contaminated at $|\tau| > |\tau_{H}|$. 

We also want to be able to associate our geometric delays with cosmological scales which we can do with
\begin{equation}
\begin{split}
k &= \sqrt{k_{\parallel}^2 + k_{\perp}^2} \\
&\approx \frac{2\pi H(z)}{\lambda(1+z)}|\tau|
\label{kmode}
\end{split}
\end{equation}
that directly relates the cosmological line of sight $k_{\parallel}$ and its perpendicular $k_{\perp}$; where $\lambda$ is the band center wavelength, $z$ is the redshift of the observation, $c$ is the speed of light, and $H(z)$ is the Hubble parameter. For our case of moving between delays and $k$ space for PAPER-64 we can approximate $k_{\perp} \approx 0$ because of the short baselines involved; this is not true for our MWA analysis which spans baselines comparable to and much longer than PAPER-64. 

\begin{figure}
	\includegraphics[width=1.\linewidth]{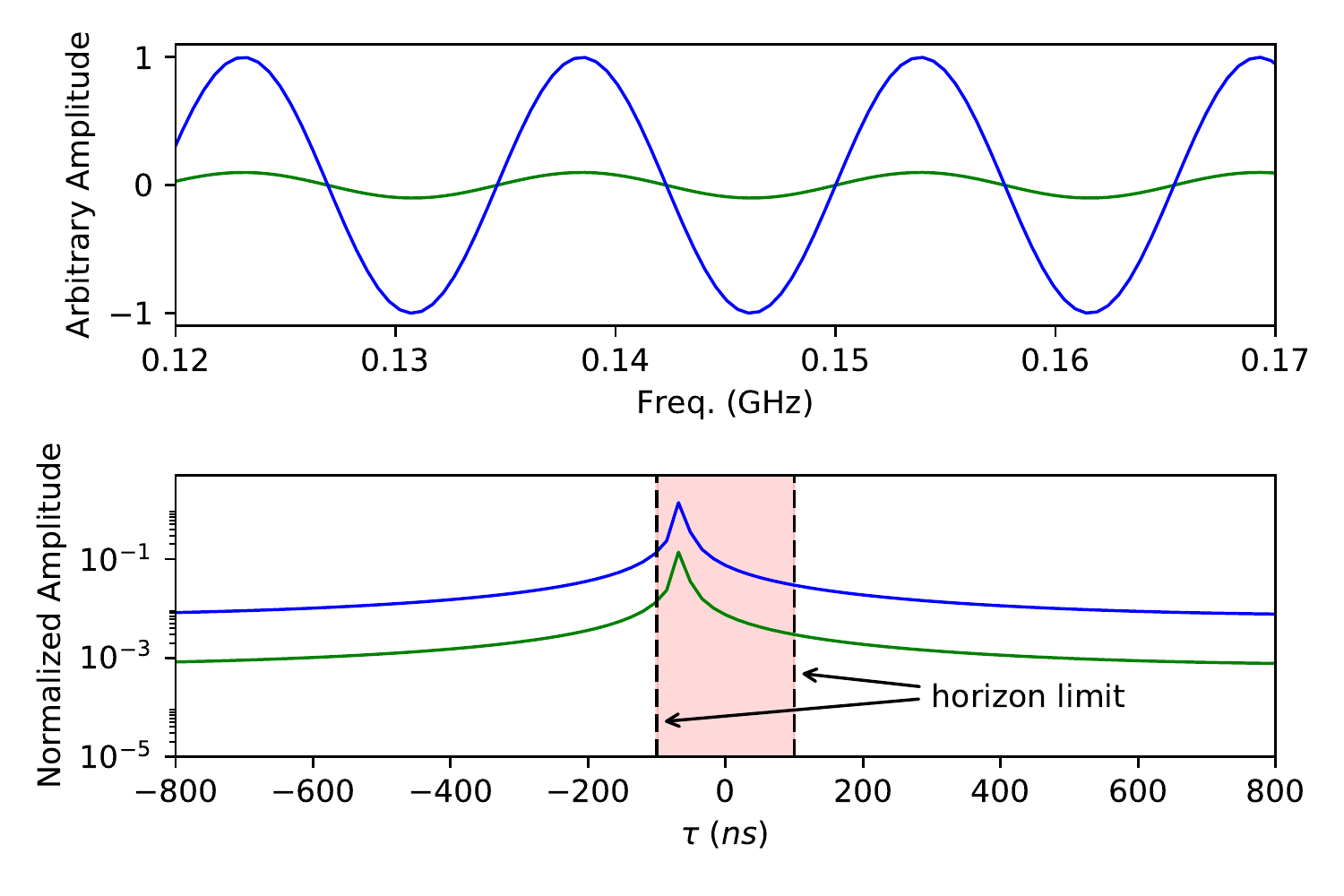}
    \caption{A toy model example of a single source near the horizon as it appears in a baseline visibility amplitude (top) and its delay transform (bottom), demonstrating how it appears near the horizon limit. The green line represents a 60$\%$ partial removal of the blue point source. Even the partial subtraction of a source can produce a non-negligible reduction in spectral leakage to delay bins outside of the horizon limit.}
    \label{fig:source_example}
\end{figure}

A toy model demonstrating an example of spectral leakage can be seen in Figure \ref{fig:source_example}, where a single point source spectrum near the horizon is shown in visibility space (top) at full intensity (blue) and reduced by 60$\%$ (green). The delay spectrum (bottom) shows the full and reduced intensity point source. Both have sidelobes that contaminate higher delays, but the reduced source has a proportionally reduced spectral leakage in delay space.  Thus, even a partial foreground subtraction can free up more sensitivity to $k$-modes where we are more likely to make a statistical EoR detection.

The approximation of foreground sources being confined within a maximum delay, $\tau_h$, gives us a natural motivation to filter out low delays. This filtering removes the foregrounds that occupy low delay modes, and in turn will reduce spectral leakage of foreground power into modes beyond $\tau_h$. 
However, due to discontinuities in the visibility sampling function (e.g. missing data from radio frequency interference (RFI) and a finite band i.e. convolution with a top-hat window function), foreground power is scattered to high delays and thus a simple delay-space filter is insufficient for mitigating spectral leakage \citep{Parsons2009,Parsons2012,Parsons2014FG}. The delay filtering in the PAPER-64 power spectrum pipeline uses the application of a per baseline and per time wide-band deconvolution filter that iteratively fits components in delay space to remove foregrounds. We describe how the wide-band delay filter is implemented here in a succinct manner so as to motivate our foreground subtraction approach for mitigating any spectral leakage from this process.  A more in-depth explanation of the filtering approach is outlined in \citet{Parsons2012}.

The deconvolution process works in delay space.  Since the sampling function in frequency is known, one can calculate the delay-space convolving kernel created by the sampling.
The algorithm then simply finds the brightest delay mode and subtracts a fraction of its power, accounting for the convolving kernel that correlates it with all other delay modes.\footnote{In previous publications, this filtering has been referred to as CLEAN or a variant thereof, leading to some confusion.  This iterative deconvolution process \emph{is} based on the CLEAN algorithm as outlined in \cite{Hogbom1974}, but there is no imaging or other steps associated with typical radio astronomy CLEAN algorithms. The PAPER wide-band delay filtering is similar only in the iterative process of deconvolving the maximal point in 1D delay space.}
The range of modes that are fit in delay space can be restricted to be within the horizon limit (or, typically, a small buffer of 15\, ns beyond the horizon limit) so that no signal is removed from the higher delays.
This process is iterated over until a predetermined foreground residual threshold is met. This threshold is determined by the ratio of the initial amplitude to the final amplitude, which for the purpose of demonstrating this technique we use $10^{-9}$. We choose $10^{-9}$ because it provides the minimum amount of worst-case signal loss ($<$5\% signal loss) at the modes nearest the horizon \citep{Ali2015} and deconvolves the maximum amount of the sampling function while still converging. Decreasing the order of this threshold or using a more sparse sampling/windowing function can lead to the algorithm's inability to converge giving results with reduced filtering. 
We provide additional statistics in Appendix \ref{appendix:A} for understanding how both a Blackman-Harris or Top-hat window function with various thresholds affect the signal loss when applying the iterative wide-band deconvolution filter.

\section{Foreground Subtraction Methodology}
\label{section:fsmethod}
Foreground subtraction by forward modeling a source catalog to match observations. This model can come from an external catalog, or can be created from the data itself, typically through a source-finding algorithm in image space.  Because of its limited imaging capabilities and poor PSF, we do not want to use data from PAPER to create a foreground model.  Rather, we use the GaLactic and Extragalactic All-sky Murchison Widefield Array (GLEAM; \citealt{Hurley-Walker2017}), described in Section \ref{section:catalog}.  To generate and subtract model visibilities, we use the Fast Holographic Deconvolution (FHD; \citealt{Sullivan2012}) software package. The specifics of the visibilities generated for this work are described in Section \ref{section:fhdsettings}.

\subsection{Model Catalog}
\label{section:catalog}
The standard MWA GLEAM catalog contains approximately 300,000 point sources across 30,000 deg$^2$. For this work, we use a modified GLEAM catalog which includes Pictor A\footnote{We use a value of Pictor A's flux density which is 10\% higher than that from \cite{Jacobs2013}.  This factor is an approximate correction for the flux density scaling issues between the VLSS survey \citep{Cohen2007} and the VLSSR survey \citep{Lane2014} used as part of the MCMC pipeline of \cite{Jacobs2013}.} and Fornax A, both of which are treated as point sources given PAPER's resolution. 

It should also be noted that the catalog contains an observing gap from 6h 30m to 7h 30m LST to avoid the Galactic plane, which lies at the tail end of the available PAPER-64 observational data. We choose to focus on the LST range (1-6 h) where known sources were subtracted and thus ignore the observational data where gaps in the GLEAM catalog exist. PAPER also has a non-trivial response to bright sources above the maximum GLEAM declination of $+30^{\circ}$ (e.g. Cygnus A).  These sources have not been modeled or removed from the data.

\subsection{Model Visibility Parameters}
\label{section:fhdsettings}

\begin{figure*}
\centering 
\subfigure[Dirty]{\label{fig:fhdskyA}\includegraphics[width=85mm,trim={1cm 0cm 0cm 0cm},clip]{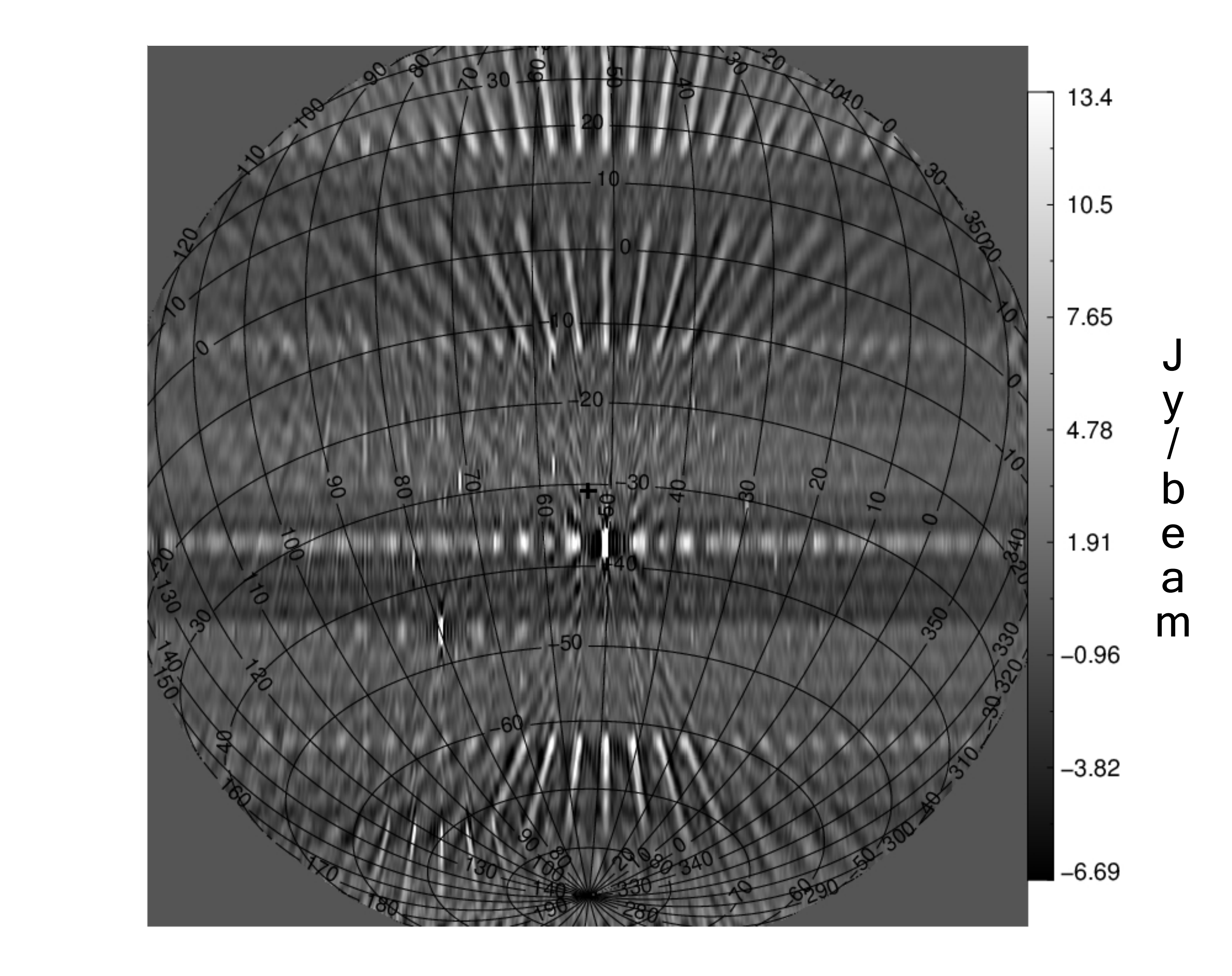}}%
~
\subfigure[Model]{\label{fig:fhdskyB}\includegraphics[width=85mm,trim={0cm 0cm 1cm 0cm},clip]{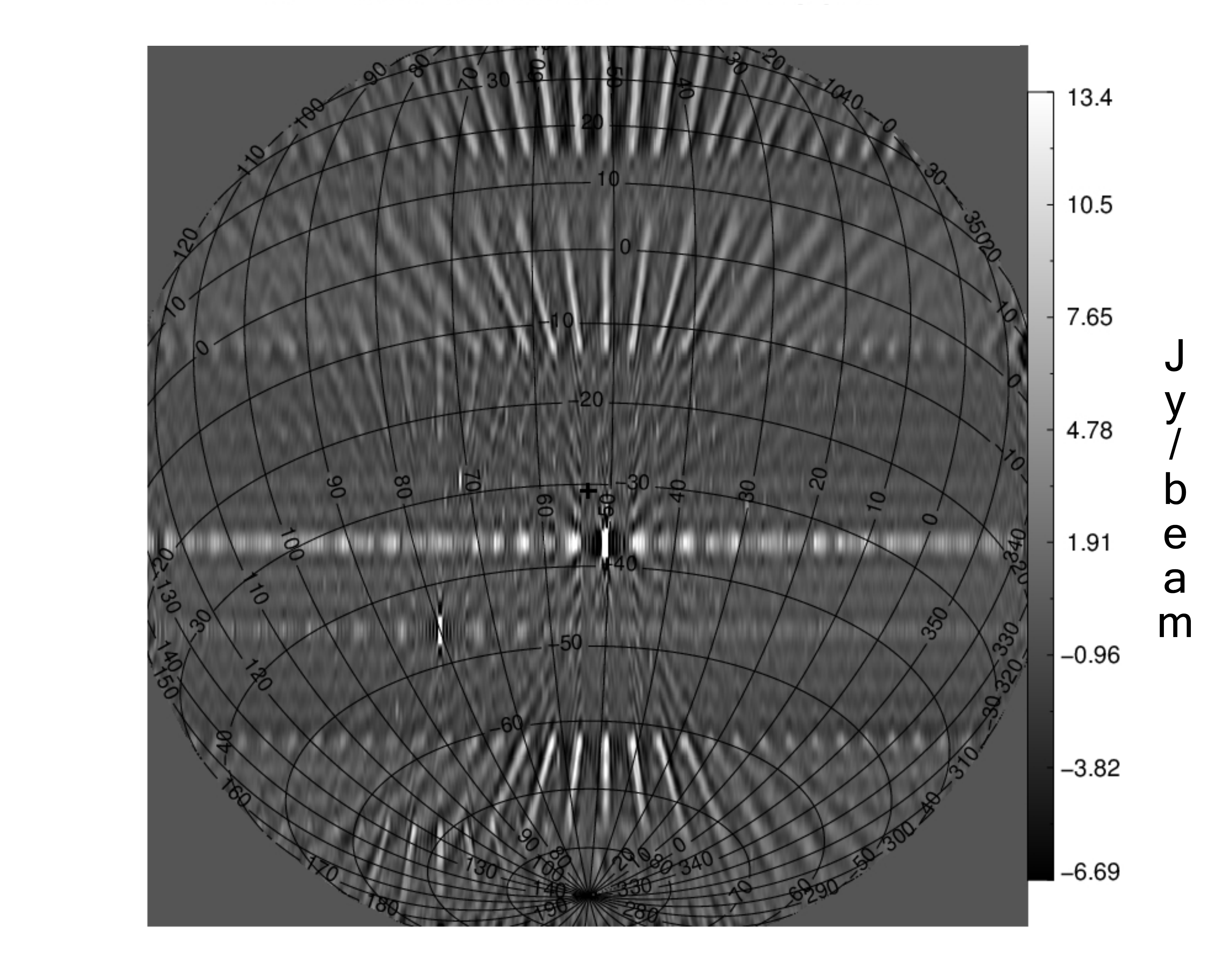}}

\subfigure[Residual]{\label{fig:fhdskyC}\includegraphics[width=85mm,trim={0cm 0cm 0cm 0cm},clip]{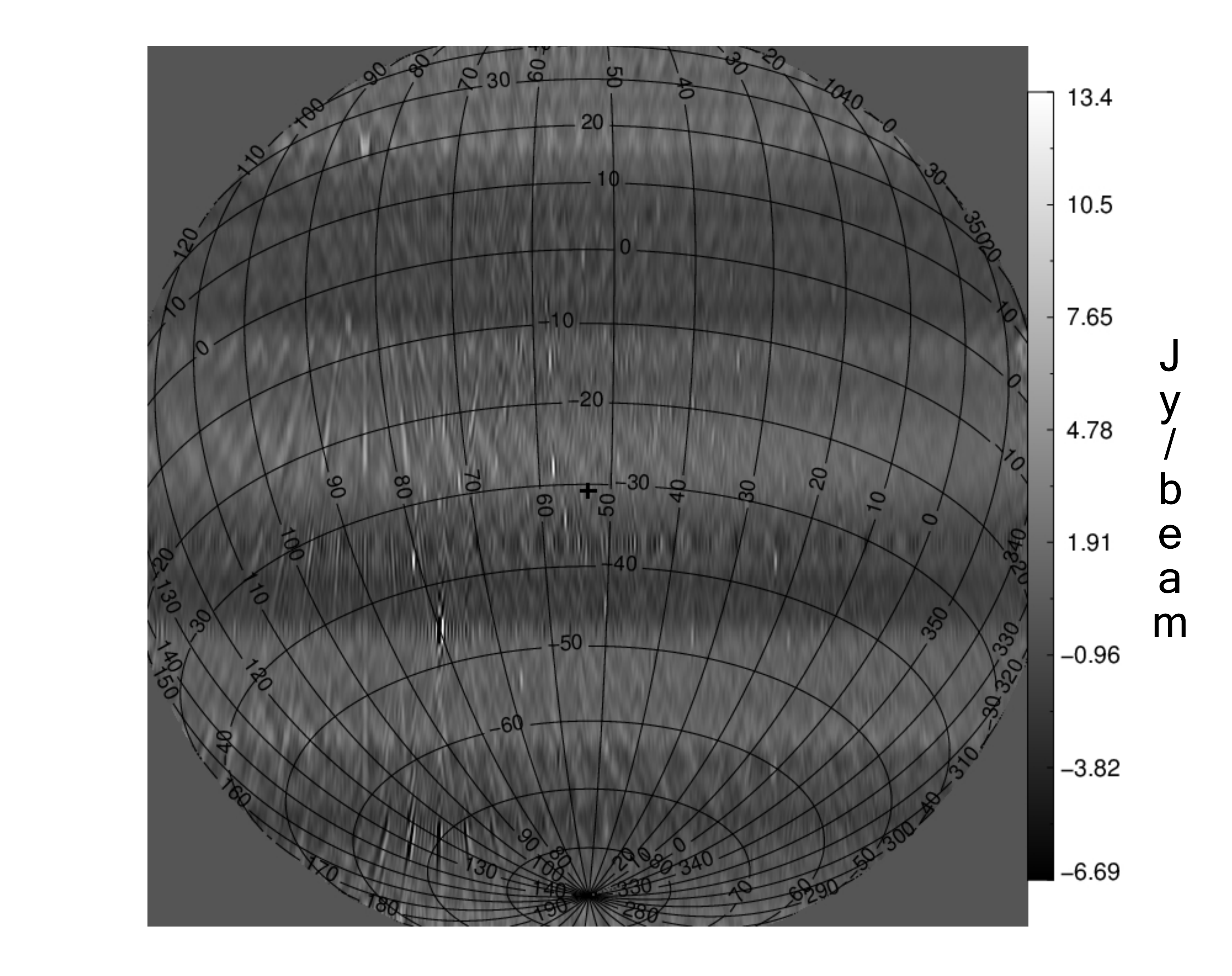}}

\caption{FHD generated Dirty (a), Model (b), and Residual (c) sky images of a single 10 minute PAPER-64 observation. The two brightest sources in the FoV are Fornax A ($3^{h} 22^{m}$,\ -37$^{\circ}$12') and Pictor A ($5^{h} 19^{m}$,\ -45$^{\circ}$46'). In the Residual sky image, it can be seen that Fornax A is removed almost entirely, while Pictor A is subtracted but to a lesser degree. The partial subtraction of Pictor A is most likely due to a poor calibration and flux model in FHD, where the brightest sources (Fornax A) dominate the calibration process. Cygnus A can be seen towards the upper-left of the Dirty and Residual sky images, but is not present in the model because it is outside of the MWA GLEAM survey field.}
\label{fig:fhdsky}
\end{figure*}

FHD has three primary functions: Simulation, Deconvolution and `Firstpass'. The Deconvolution feature allows for full deconvolution of sources using a modified CLEAN algorithm \citep{Hogbom1974,Sullivan2012}. Using FHD's full deconvolution is both time consuming and computationally expensive, and thus is not reasonable to fully deconvolve foregrounds from a 128 day dataset. We choose to use the \textit{Firstpass} feature that accepts a source catalog, a primary beam model \citep{Pober2012}, and the antenna layout which is then forward modeled to match a given observation. This reduces the time for modeling (approximately 10,000 point sources) and subtraction of a single 10 minute PAPER-64 observation to $\sim 20$ minutes using considerable computing resources \footnote{For a single observation modeled with a $uv$-plane resolution of 0.35 $\lambda$ we used 3 cores from an Intel Xeon E5-2650 v4 with 80 GB of RAM with a runtime of $\sim$ 20 minutes. Comparatively for a resolution of 0.1 $\lambda$ using the same 3 cores and 100 GB results in a runtime of $\sim$ 45 minutes}. We force all FHD foreground subtractions to model the GLEAM catalog as point sources, as PAPER-64 lacks the ability to properly resolve any of the extended sources in its FoV. An example of a single observation showing poorly resolved extended sources and the successful subtraction of Fornax A is shown in Figure \ref{fig:fhdsky}, where Fornax A ($3^{h} 22^{m}$,\ -37$^{\circ}$12') is near the center of the FoV. 

FHD's Firstpass mode also includes a calibration of the data using the model visibilities and the method described in \cite{Salvini2014}.  However, the PAPER data analyzed here is already calibrated using the methods described in \cite{Ali2015}. Hence, we allow FHD to fit a $3^{\rm{rd}}$ order polynomial to the time averaged per-frequency solutions for both phase and amplitude to correct for any gain scale or phase center discrepancies between the PAPER calibration and the GLEAM catalog. We calibrate and model 10,000 point sources of fixed flux density ($>$ 1 Jy) and position from the GLEAM catalog. Point sources are modeled in FHD as having a flat spectrum.

\subsection{Excess High Delay Power in FHD Models}
\label{sec:excess-power}
When modeling sources in FHD, each GLEAM point source is modeled as flat across frequency. In the absence of numerical errors, we would therefore expect that the only power that is pushed to higher delay modes in the delay transform is due to spectral leakage from a finite band and from the primary beam's spectral structure. In principal, FHD can simulate precise and realistic spectral responses when computational resources are abundant, however in this analysis approximations must be made. Numerical artifacts from these approximations in the visibility simulation process can introduce spurious spectral structure in the model visibilities. One particular source of artificial spectral structure in the algorithm used by FHD comes from its discrete sampling of the $uv$ plane at a finite resolution. In this work, to balance the needs of precision and computational efficiency with a 128 day data-set, we model the $uv$ plane with a pixel size of 0.35 $\lambda$ on a side. This does not completely alleviate the numerical artifacts in the source spectrum; instead, in delay space, we find an aliasing floor beyond the horizon that is $\sim 3$ order of magnitude smaller than the foreground amplitude, but still several orders of magnitude greater than a 21cm EoR signal. When these model visibilities are subtracted from our PAPER visibilities, this high delay power is transfered into our residual visibilities.  With the sensitivity of this dataset, this spurious spectral structure in the foreground model becomes the limiting factor in the analysis and sets a floor to any potential 21cm signal limit.

To make progress, we low pass filter our model visibilities using the iterative deconvolution filter with limits of $|\tau| \geq 90$ ns (for 30m baselines). This removes a minimal amount of our model within the horizon limit while giving us the dynamic range to ensure that we are not adding power that would contaminate the EoR signal at high k-modes and near the horizon. We present a more in-depth analysis into this source of spurious spectral structure and the filtering we perform on model visibilities in Appendix \ref{appendix:B}.

This filter indiscriminately removes high delay power from the model visibilities, including potential effects from the intrinsic spectral structure in the foregrounds or the instrument response.  In this work, the foregrounds are modeled as flat spectrum, and so possess no intrinsic high delay power; however, we do use the full frequency-dependent PAPER beam model in creating the model visibilities.  Ultimately, we would like to include the full frequency-dependent model of the instrument and a frequency-dependent sky model in this kind of analysis. At present, however, the spurious structure limits our analysis and filtering is a necessary step to reach sensitivities relevant to our analysis. We further discuss the implications of this analysis choice in Section \ref{section:conclusions}.

\section{Data and Preprocessing}
\label{section:data}
\subsection{PAPER}

The drift-scan PAPER observations used for this analysis originate from the November 2012 - March 2013 observing season of 128 days using the 64 element dual-polarization PAPER array. The dataset is compressed in both time and frequency by averaging from 10.7 s time integrations to 42.9 s and from frequency channel widths of 97.7 kHz to 492.6 kHz using the delay/delay-rate filtering scheme described in \cite{Parsons2014FG}. We identify and flag RFI at the 6$\sigma$ level prior to calibration. Observations are then redundantly calibrated, also known as OMNICAL calibrated \citep{Zheng2014}, and absolute calibrated. Additional details of this dataset can be found in \cite{Ali2015}. Following these initial pre-processing steps we begin the foreground filtering and filtering+subtraction pipeline analysis as laid out in Figure \ref{flowchart:pspecpipeline}. It should be noted that these visibilities undergo an additional calibration in FHD (step B), but after subtraction we return each visibility to their original OMNICAL calibration. In doing this, we should be mitigating any flux density scaling discrepancies or coordinate/pointing errors. 

\subsection{MWA}
\label{sec:mwadata}
The MWA observations used are 2.6 hours from EoR0 ((RA,Dec: $0^{h}$,-27$^{\circ}$) which is a region of the sky with minimal Galactic emissions \citep{Beardsley2016}) where each visibility has a time integration of 2 s as opposed to the 42.9 s time averaged integrations of PAPER. The MWA Phase I EoR observations \citep{Jacobs2016,Trott2016FG,Pober2016} have an instantaneous bandwidth of 167-197 MHz across 384 channels. The narrow-band EoR redshift observations are taken over 95 channels (7.68 MHz) which correspond to a $\Delta z \approx 0.3$ for most measurements. The pipeline for comparing filtering vs. filtering+subtraction for the MWA data is only similar to PAPER-64 up to step D and we omit the RFI flagging after the wide-band delay filter (step C).

\section{Implementation on PAPER Data}
\label{section:pipeline}
\tikzstyle{decision} = [diamond, draw, fill=blue!20, 
    text width=3.5em, text badly centered, node distance=2cm, inner sep=0pt]
\tikzstyle{block} = [rectangle, draw, fill=yellow!20, 
    text width=5em, text centered, minimum height=3em]
\tikzstyle{largeblock} = [rectangle, draw, fill=orange!20, 
    text width=20em, text centered, rounded corners, minimum height=3em,minimum width=24em]
\tikzstyle{node} = [rectangle, node distance=1cm,
    text width=10em, text centered, rounded corners, minimum height=0em]
\tikzstyle{line} = [draw, -latex']
\tikzstyle{cloud} = [draw, ellipse,fill=red!20, node distance=3cm,
    minimum height=2em]
 \tikzstyle{filter}=[rectangle, draw, fill=red!20, 
    text width=5em, text centered, minimum height=3em, minimum size = 10mm]
 \tikzstyle{noloss}=[rectangle, draw, fill=yellow!20, 
    text width=5em, text centered, minimum height=3em, minimum size = 10mm]
 \tikzstyle{evenodd}=[rectangle, draw, fill=green!20, 
    text width=5em, text centered, minimum height=3em, minimum size = 10mm]
 \tikzstyle{image} = [rectangle, fill=white!20, 
    text width=3.5em, text centered, node distance=2cm, inner sep=0pt]

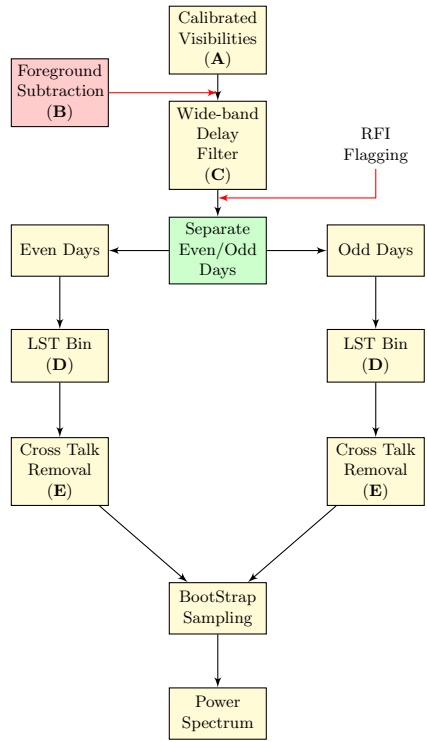
\begin{figure}[!t]
\begin{center}
\begin{tikzpicture}[node distance = 3cm, auto,scale=0.7, every node/.style={transform shape}][t!]

    \node [block] (init) {Calibrated Visibilities (\textbf{A})};
    \node [filter, left of=init,yshift=-1cm] (fgsub) {Foreground \\ Subtraction (\textbf{B})};
    \node [block,below of=init,yshift=1cm] (Set) {Wide-band Delay Filter \\ (\textbf{C})};
    \node [evenodd, below of=Set,yshift=1cm] (sep) {Separate Even/Odd Days};
     \node [node,below of=Set,xshift=3cm,yshift=1cm] (RFI) {RFI \\ Flagging};

    \node [block, left of=sep] (even) {Even Days};
     \node [noloss, below of=even,yshift=1cm] (sepnode1) {LST Bin \\ (\textbf{D})};
    \node [block, right of=sep] (odd) {Odd Days};
    \node [noloss, below of=odd,yshift=1cm] (sepnode2) {LST Bin \\ (\textbf{D})};
    \node [block,below of=sepnode1,yshift=0.8cm] (frf1) {Cross Talk Removal \\ (\textbf{E})};
        \node [block,below of=sepnode2,yshift=0.8cm] (frf2) {Cross Talk Removal \\ (\textbf{E})};
  \node [block, below of=sep,yshift=-3.8cm] (BSsample) {BootStrap Sampling};
  \node [block, below of=BSsample,yshift=1.cm] (PSPEC) {Power Spectrum};

      \path [line] (init) -- (Set) ;
      \path [line] (Set) -- (sep);
      \path [line] (sep) -- (even);
      \path [line, red] (fgsub) -- ($(init)!.5!(Set)$);
      \path [line, red] (RFI) |- ($(Set)!.5!(sep)$);
      \path [line] (even) -- (sepnode1);
      \path [line] (sepnode1) -- (frf1);
      \path [line] (frf1) -- (BSsample);
      \path [line] (odd) -- (sepnode2);
      \path [line] (sep) -- (odd);
      \path [line] (sepnode2) -- (frf2);
      \path [line] (frf2) -- (BSsample);
      \path [line] (BSsample) -- (PSPEC);

\end{tikzpicture}
\end{center}
\caption{\label{flowchart:pspecpipeline}Abridged PAPER power spectrum pipeline used in processing visibilities for EoR measurements. After calibration (step A; described in Section \ref{section:data}), there is the option to subtract model foregrounds (step B).  After the wide-band filtering (step C; described in Section \ref{section:wideband}), additional weak RFI is removed by flagging data with $> 3 \sigma$ deviation. Visibilities are split into even/odd days and averaged in local sidereal time (LST; step D). We cross multiply the even/odd LST binned visibilities to obtain an unbiased power spectrum estimator. Cross-talk removal (step E) is performed by using the full 8.5 hrs of observations to ensure at least one full fringe period has elapsed. The PAPER power spectrum analysis from \cite{Ali2015} additionally uses a signal loss correction loop and inverse covariance weighting; these steps are not included here as they do not contribute to the understanding of filtering+subtraction. The letters labeling each step correspond to the steps in Figure \ref{fig:PSPECFlow}.}
\end{figure}

\begin{figure*}
	\includegraphics[width=1.\textwidth]{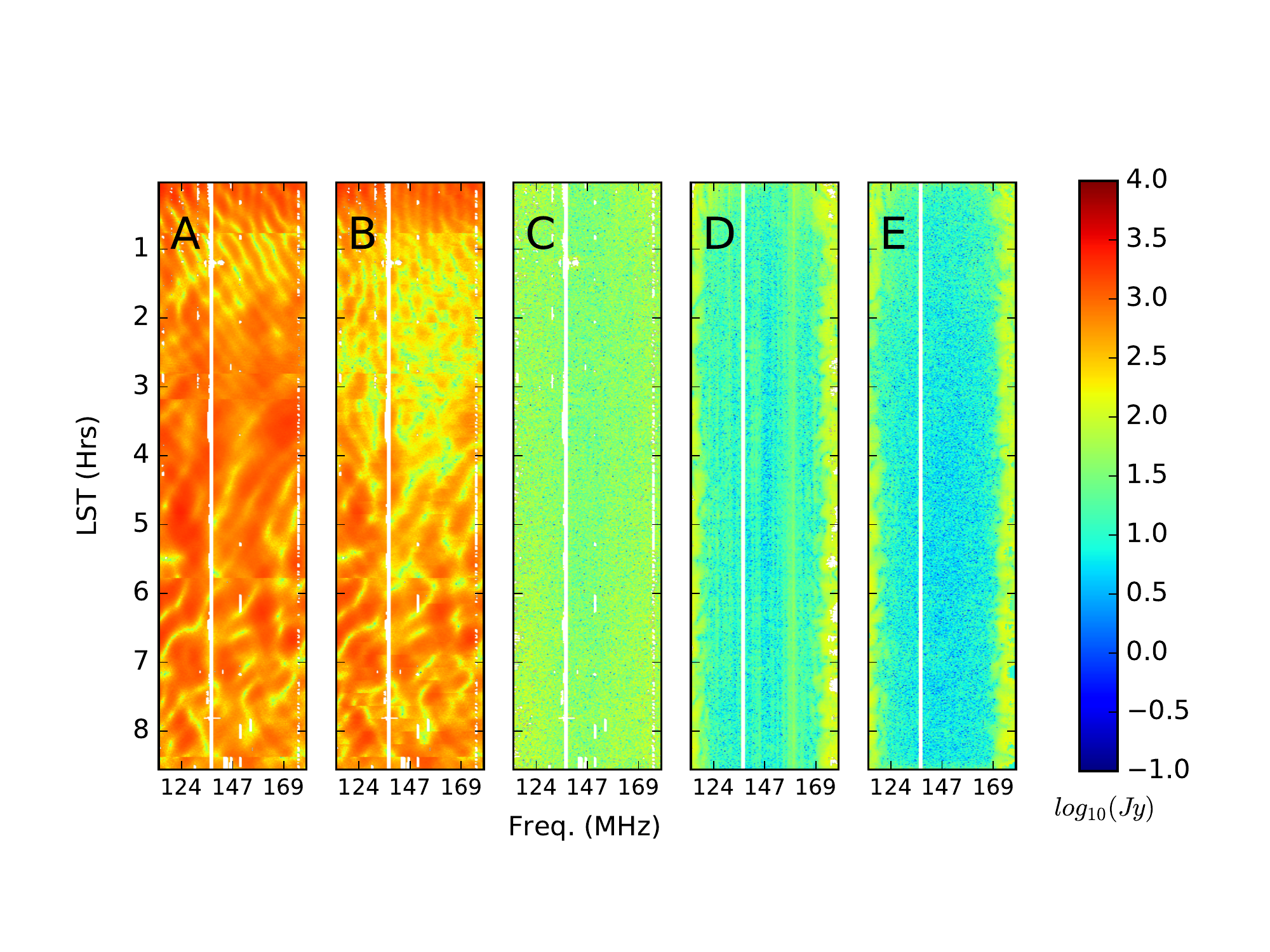}
    \caption{Sequential foreground removal steps of a fiducial 30\, m PAPER-64 baseline. The steps are (\textbf{A}) OMNICAL calibrated baseline visibility, (\textbf{B}) foreground subtraction, (\textbf{C}) foreground filtering, (\textbf{D}) LST binning, and (\textbf{E}) cross-talk removal. Foreground subtraction can be seen to reduce power primarily due to Fornax A around 1-4 hrs LST in (\textbf{B}), and additional foreground power is removed from smaller transiting sources up until the GLEAM catalog cutoff at 6h 30m LST. The end result of applying the PAPER-64 power spectrum pipeline techniques results in $\sim$ 4 orders of magnitude in amplitude of foreground suppression. A diagrammatic overview of how the pipeline progresses by step is shown in Figure \ref{flowchart:pspecpipeline}. The discontinuities seen in LST in both A \& B steps originate from gain variations due to applying the OMINICAL calibration on single 10 minute observations.}
\label{fig:PSPECFlow}
\end{figure*}

The analysis focus for this work begins after the calibration steps described in Section \ref{section:data}. We emphasize only the critical steps required for the application of the combined foreground subtraction and filtering approach; a full description of the PAPER-64 analysis pipeline can be found in \cite{Ali2015}. An abridged outline of the power spectrum pipeline can be seen in Figure \ref{flowchart:pspecpipeline}, which shows the placement of the foreground subtraction step we add to the analysis. We choose to forgo the fringe rate filtering and the inverse covariance weighting at the end of the traditional PAPER-64 power spectrum pipeline. While these steps can improve the overall limit one can place on the 21cm signal, they obscure many of the main effects seen in our analysis and complicate interpretation of the results. However it is possible to include inverse covariance weighting, fringe rate filtering, and signal loss in the filtering+subtraction power spectrum pipeline. These additional techniques can be included by following the results from the forthcoming paper \emph{Cheng et. al.} (\textit{in prep}) which discusses issues with applying inverse covariance weighting and the signal loss involved.
To determine the effect of foreground subtraction on the analysis, we split our pipeline into the traditional PAPER-64 power spectrum pipeline, which we refer to as \textbf{filtering}, and one with FHD foreground subtraction included, referred to as \textbf{filtering+subtraction}. Figure \ref{fig:PSPECFlow} shows the effect each of these foreground removal steps has on the visibility amplitudes of 8.5 hrs of data from a 30\, m baseline. The first three steps (A,B,C) involve no averaging over days of observation; however, in panel D LST binning is used to average over 64 days of observation (while the total data set is $\sim$128 days, even and odd days are averaged separately). Note that omitting foreground subtraction (step B) results in plots C through E which are visually indistinguishable from the version shown. The white regions in Figure \ref{fig:PSPECFlow} are missing data due to flagging intermittent RFI which are recovered in step D due to LST binning. The RFI flagging that is consistent among all steps centered at 137 MHz with no recovery from LST binning is the ORBCOMM satellite system which is persistent across all observations and confined to $\sim$4 frequency channels.

Depending upon the RFI strategy employed, frequency/time samples can be falsely identified as containing RFI, which is the case when flagging between steps C to D. The wide-band delay filtering decreases the narrow-band variance in the band center due to the Blackman-Harris window function; however, the wide-band variance remains comparatively large. This means that a reduction in the false-positive RFI identification rate can be achieved from either reducing the wide-band power (foreground subtraction) or using a different window function in the filtering step prior to RFI flagging. We address additional benefits of using a different window function in the filtering step and how it affects the power spectrum in Section \ref{sec:windowfunctions}. 

In the final pre-processing step prior to the power spectrum estimate we apply cross-talk removal to all visibilities. We average per frequency over the entire 8.5 hrs of visibilities which is then subtracted from each observation. By doing this we are able to account for the offset in the visibility complex plane due to cross-talk that becomes apparent after one full fringe period in our 30m baselines.

For determining the statistical significance of the filtering+subtraction technique we empirically estimate errors on the PAPER-64 power spectra in Section \ref{sec:bandcenter} by bootstrapping. This bootstrapping is done in accordance with modifications suggested in \emph{Cheng et. al.} (\textit{in prep}). We use 400 bootstrap samples taken from three redundant 30m baseline types and over LST. To properly account for the error of differenced power spectra in Sections \ref{sec:bandcenter}, \ref{sec:windowfunctions}, and \ref{sec:buffer}, we bootstrap over the distribution of differenced samples rather than use the combined error from each independent power spectrum. The differenced power spectra are created in parallel to make certain that the same random sample from LST and baseline are chosen.

\subsection{Effect of Foreground Subtraction}
\label{sec:bandcenter}

\begin{figure*}
\centering 
\subfigure[Power Spectra]{\label{fig:PowerSpectraMid}\includegraphics[width=9.cm,trim={0cm 0cm 1.cm 1cm},clip]{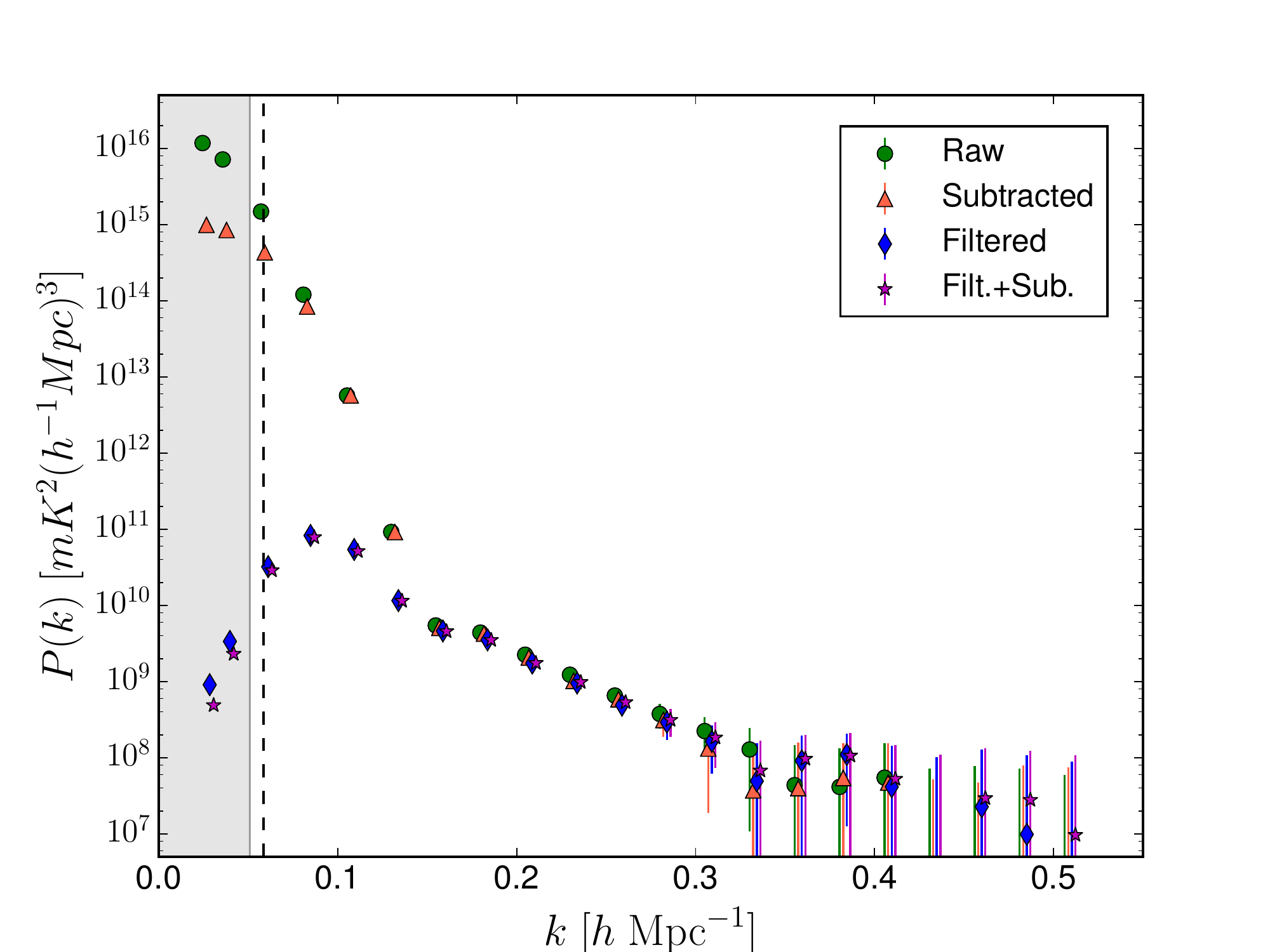}}%
~
\subfigure[Differenced Power Spectra]{\label{fig:PowerSpectraMidDiff}\includegraphics[width=9.cm,trim={0cm 0cm 1.cm 1cm},clip]{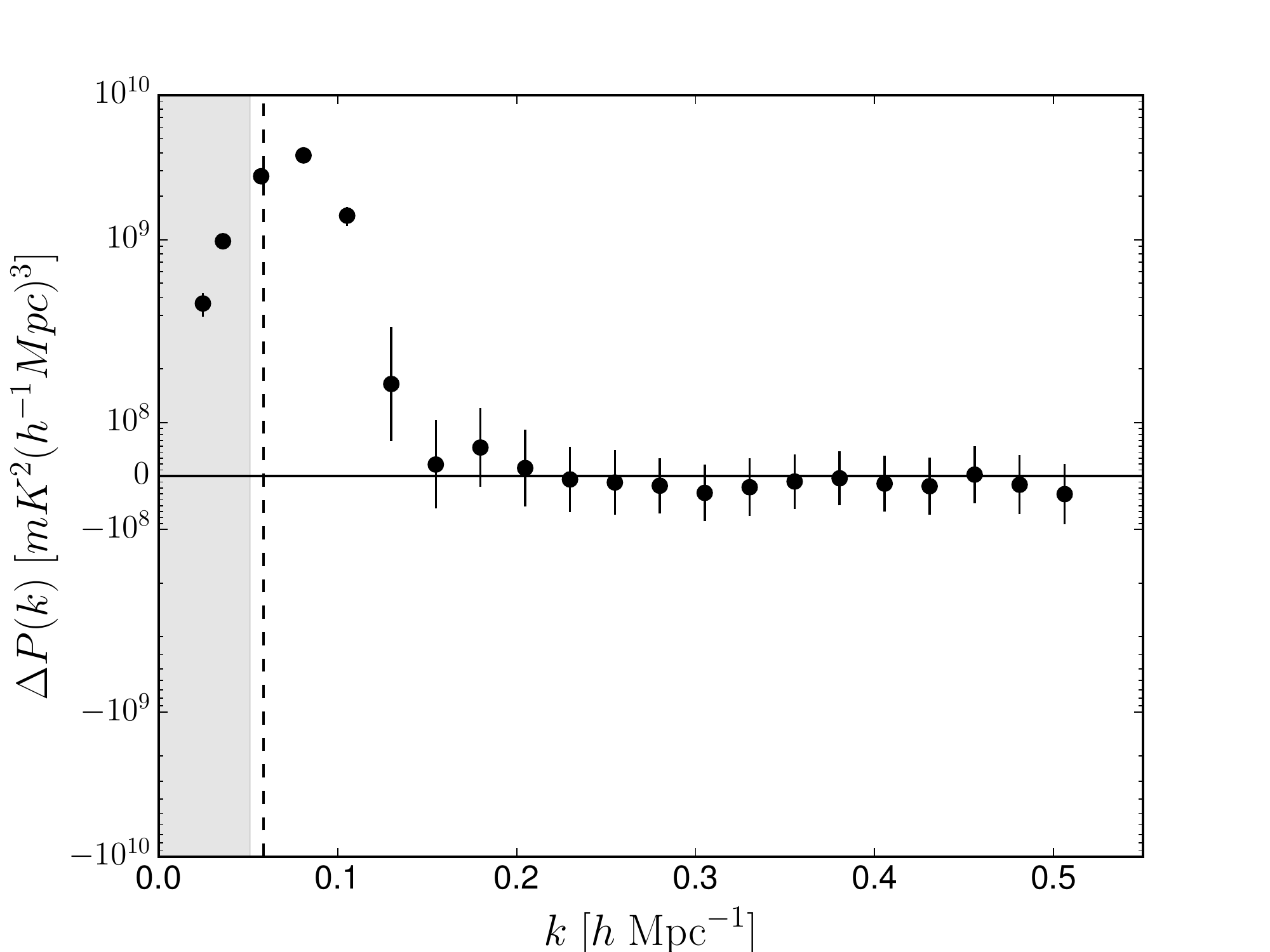}}%
~
\caption{(a). Resulting power spectra from the 128 day LST binned PAPER-64 dataset at band center (z=8.4) focused on 1-6 hrs LST, where the shaded region represents the horizon limit for a 30m baseline and the broken vertical line is a 15 ns filtering extension. Power spectra correspond to each step of foreground removal: the raw PAPER-64 power spectrum (green), foreground subtracted (orange), foreground filtered (blue), and our filtered+subtracted technique (purple). (b). The differenced filtered and filtered+subtracted power spectra showing where we see improvements in sensitivity; more positive means greater sensitivity improvement due to reduction in spectral leakage. The first few modes nearest the horizon ($> k \approx 0.06$) show a statistically significant ($2\sigma$) improvement through the implementation of filtering and subtraction.}
\end{figure*}

We use the PAPER-64 power spectrum pipeline from \citet{Ali2015} as our 
point of comparison for determining how applying foreground subtraction 
prior to filtering can increase our sensitivity. The significance of the 
improved sensitivities are estimated by bootstrap sampling over our
distribution of power spectra and differenced power spectra. 
While the foreground filter is performed across the entire 100\, MHz 
PAPER band (hence the name ``wide-band filter"), cosmological power 
spectra are restricted to a narrower range of frequencies to limit 
redshift evolution.  In this analysis, we use 20 MHz of data for each 
cosmological power spectrum, but use a Blackman-Harris window function 
in this ``narrow band'' delay transform, giving an effective bandwidth 
of 10 MHz. 

Power spectra from the band center at 150\, MHz ($z = 8.4$) 
can be seen in Figure \ref{fig:PowerSpectraMid}, where we compare each successive technique for foreground removal: no foreground mitigation 
(green), foreground subtraction (red), foreground filtering (blue), and 
filtering+subtraction (purple). The shaded gray region corresponds to 
modes inside the horizon, and the dashed black line corresponds to a 
delay of 15\, ns beyond the horizon; the wide-band deconvolution is only 
allowed to fit modes with delays less than this value. We see that 
subtraction drops the foreground power by an order of magnitude within 
the horizon limit, with marginal improvements at higher delays.  
(Because of the Blackman-Harris window function in the narrow band delay 
transform, adjacent $k$ modes in the power spectrum are highly 
correlated, and so the improvements in the first mode beyond the horizon 
should be interpreted with caution.)  The foreground filtering removes a 
very large fraction of power within the horizon limit, but also leads to 
significant improvements in the first several modes beyond the horizon.  This behavior is indicative of reduced spectral leakage from modes 
within the horizon.

The power spectrum difference between filtering and filtering+subtraction can be seen in Figure \ref{fig:PowerSpectraMidDiff}, which 
shows a statistically significant ($\gtrsim2\sigma$) improvement of 
modes within and near the horizon. At the most basic level, this result 
is expected. The model-based foreground subtraction has made the job of 
the wide-band delay filter easier: there is less power it needs to 
remove.  In particular, when foregrounds are subtracted from the raw 
visibilities, we reduce the amount of power that rings off of sharp 
discontinuities in the frequency sampling function when taking the delay 
transform. In the PAPER data set, the ORBCOMM satellite band centered at 137.5\, MHz spans several channels and is always flagged. This creates a very sharp discontinuity that leads to strong ringing of foreground power throughout visibility space. Because the frequency sampling function is known, the iterative wide-band filter can, in principle, remove this scattered power itself. But, the deconvolution algorithm is not perfect, and it appears that reducing the total amount of foreground power available to scatter in a delay transform leads to better performance of the wide-band deconvolution filter. Put another way, the wide-band foreground filter has an effectively fixed dynamic range that it can achieve between the brightest foreground contaminated modes and the limits in the EoR window.  Even though the model visibilities contained no intrinsic power in the EoR window, reducing the power in the brightest foreground contaminated modes improves the end result because we are limited by the dynamic range of the wide-band filter.

\subsection{Performance Near Band Edges}
\label{sec:windowfunctions}

\begin{figure*}
\centering 
\subfigure[Low Band (z = 10.4)]
{\label{fig:3PlotLow}\includegraphics[width=93mm,trim={0cm 0cm 0cm 1cm},clip]{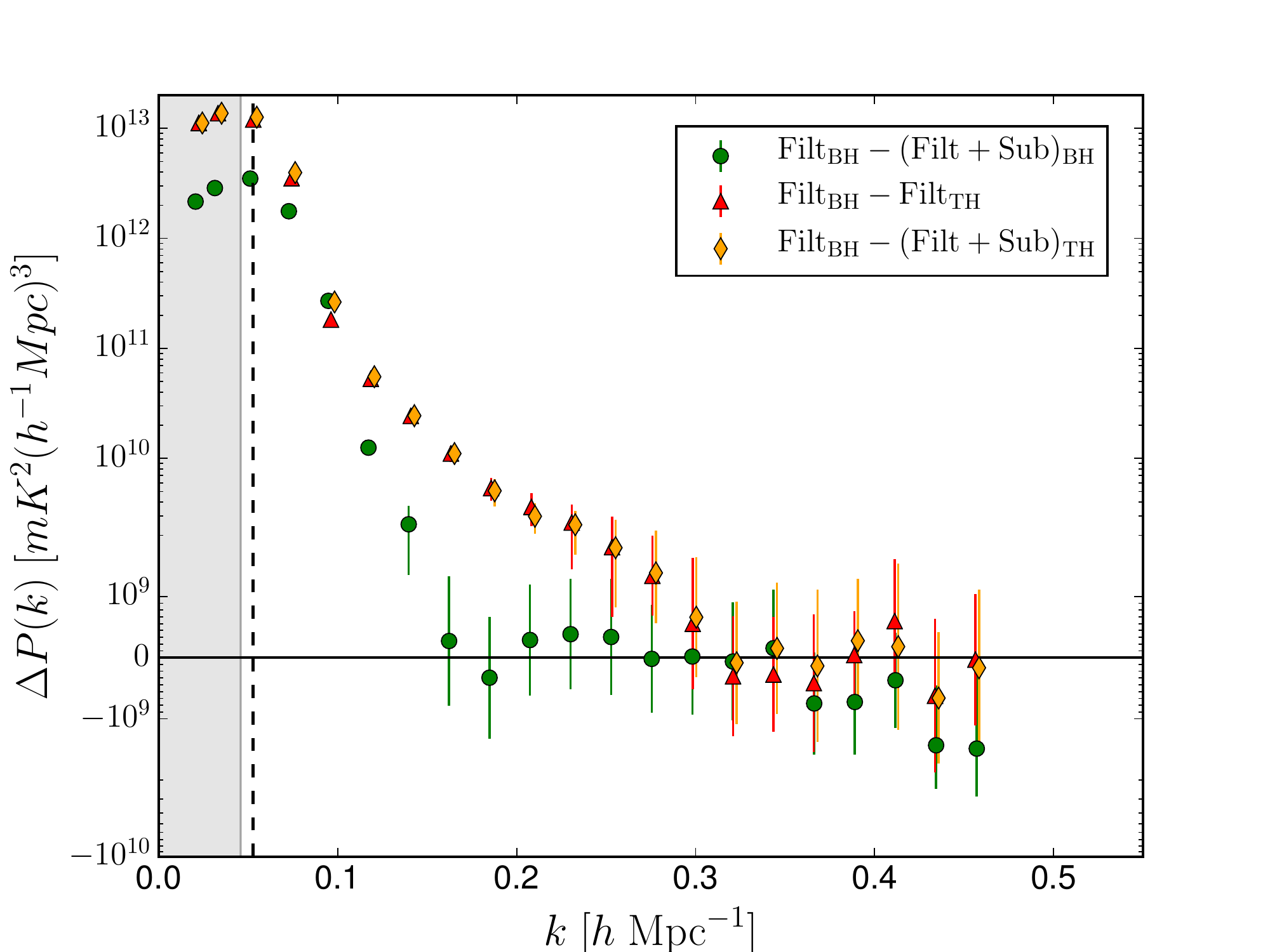}}%
~
\subfigure[Mid Band (z = 8.4)]
{\label{fig:3PlotMid}\includegraphics[width=93mm,trim={0cm 0cm 0cm 1cm},clip]{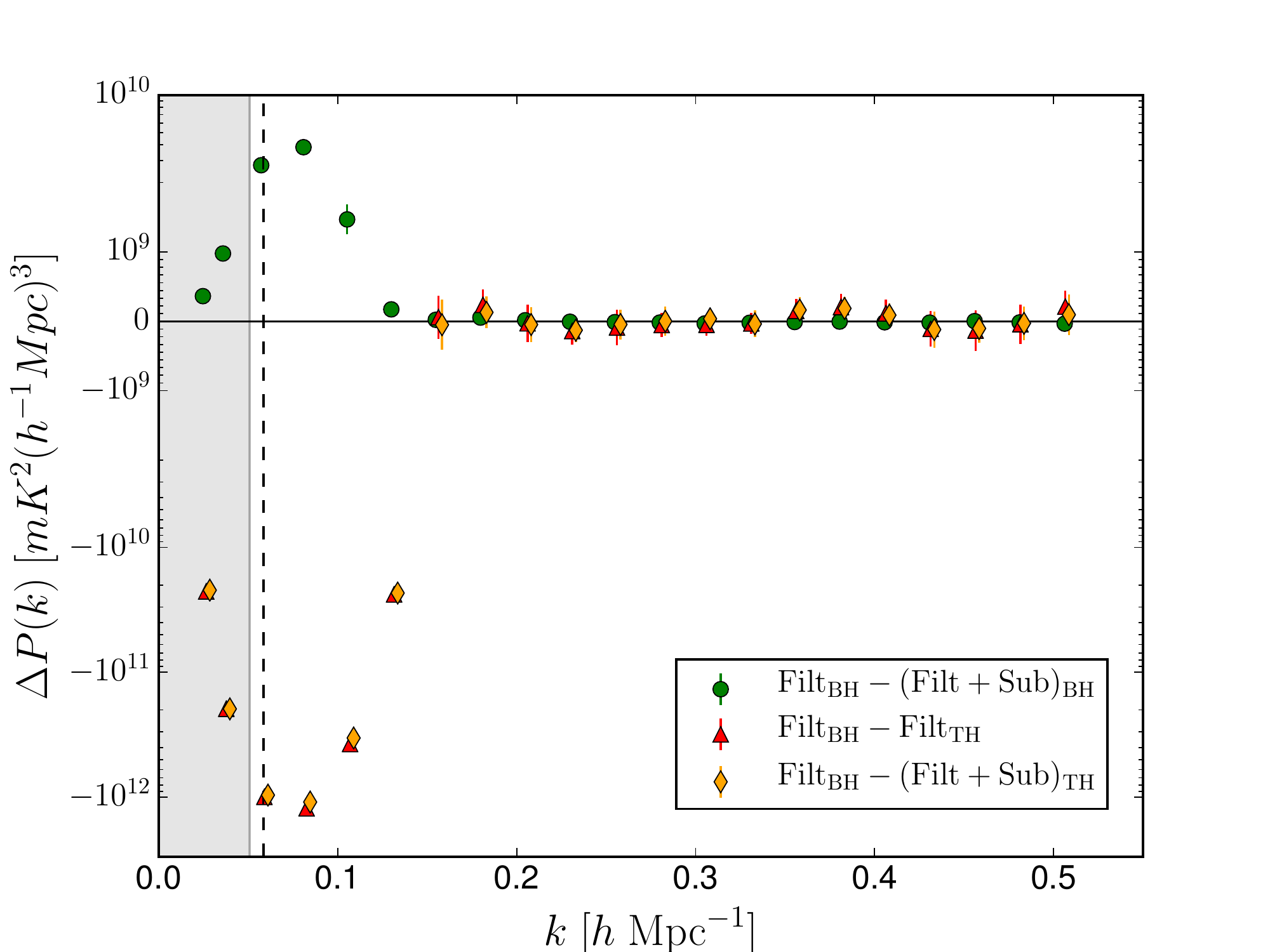}}

\subfigure[High Band (z = 7.4)]{\label{fig:3PlotHigh}\includegraphics[width=93mm,trim={0cm 0cm 0cm 1cm},clip]{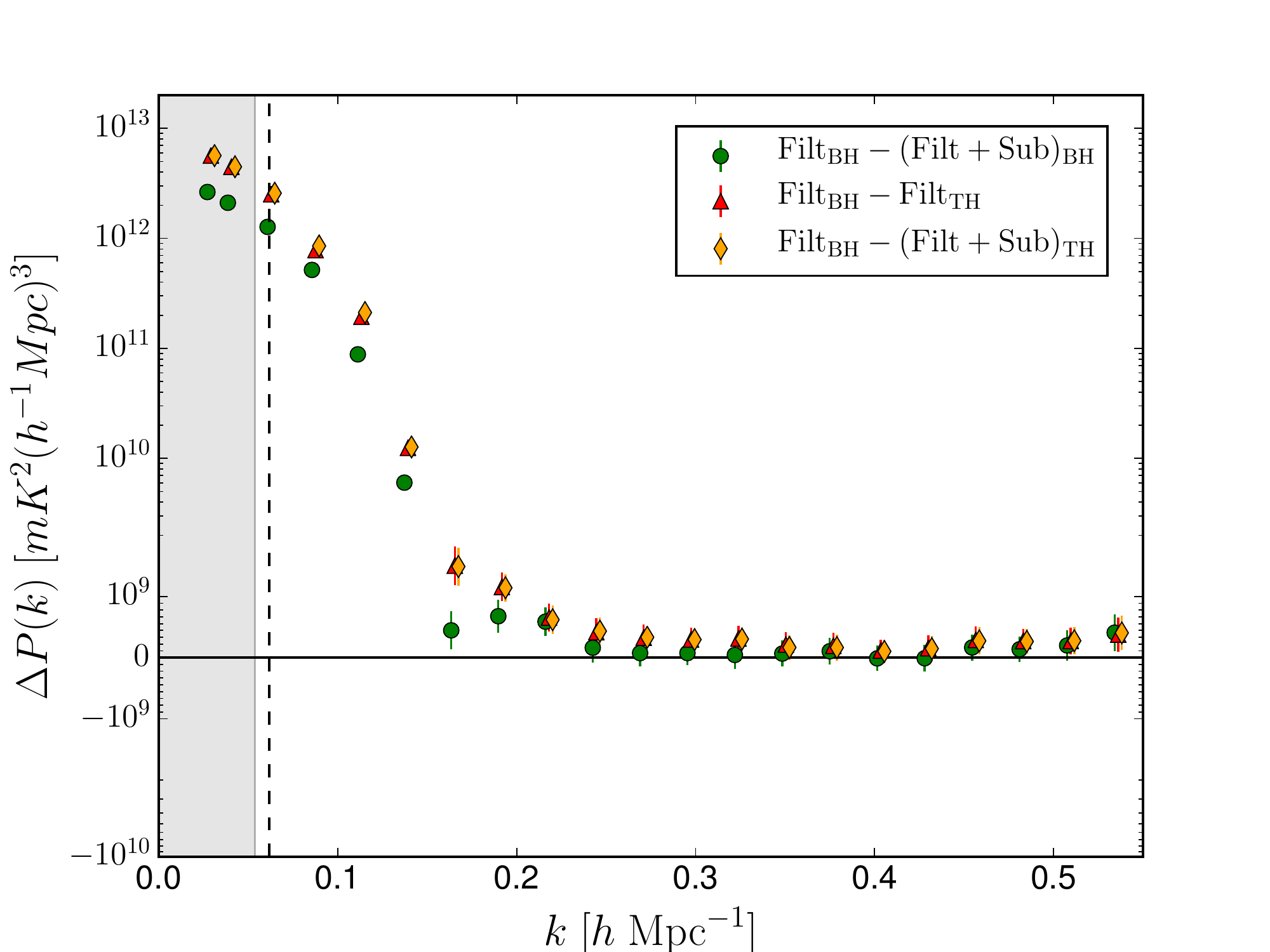}}

\caption{The power spectrum difference between filtering and filtering+subtraction at different band locations comparing the effectiveness of window function choice on power removed during the wide-band filtering step, where again more positive shows an improvement by both filtering and subtracting foregrounds. We compare the standard Blackman-Harris (BH) to the Top-hat (TH) window function. We should expect that the Top-hat window function gives us the maximum amount of foreground removal at band edges because of the equal weighting across the entire band. A symmetric improvement about band center is shown in (a) \& (c) where the switch from BH to TH window function increases the reduction in spectral leakage in the low (high) band by $\sim$ 145\% (113\%) and with subtraction $\sim$ 169\% (125\%) at $k\approx0.11 \ \rm hMpc^{-1}$. The band center plot (b) demonstrates that the BH window function is the ideal choice regardless of filtering or filtering+subtraction when compared to the TH window.}
\label{fig:threezdiff}
\end{figure*}

One of the strengths of the PAPER experiment is its wide instantaneous bandwidth which covers from 100 to 200 MHz ($z \sim 6 - 13$). The data analyzed in Section \ref{sec:bandcenter} were centered around 150 MHz, where past analyses have shown the wide-band foreground filter (using a Blackman-Harris window function) to be most effective \citep{Jacobs2015}. In this section, we demonstrate the effectiveness of filtering+subtraction nearer to the edges of PAPER's frequency band with power spectra at 124\, and 169\, MHz (corresponding to $z = 10.4$ and 7.4, respectively).

The green points in Figure \ref{fig:threezdiff} show the difference in power spectra made using filtering and filtering+subtraction at all three redshifts: $z = $ 10.4 (Figure \ref{fig:3PlotLow}), 8.4 (Figure \ref{fig:3PlotMid}), and 7.4 (Figure \ref{fig:3PlotHigh}).  (Note that the green points in Figure \ref{fig:3PlotMid} are the same data shown in Figure \ref{fig:PowerSpectraMidDiff}.) These data show that filtering+subtraction removes significantly more power from the first few modes outside the horizon than filtering alone.  Compared with the $z=8.4$ results in Figure \ref{fig:PowerSpectraMid}, we see that filtering+subtraction removes one to two more orders of magnitude of foreground power than it does at the band center. This increased effectiveness of our technique at the band edges comes more from foreground subtraction itself than any interplay between the subtraction and the filtering, which drove the improvements at band center. This effect is primarily because the Blackman-Harris window in the wide-band filtering step specifically down-weights the band edges. Therefore, we expect the wide-band filter to perform comparatively poorly at the band edges, as was seen in \cite{Jacobs2015}. Subtraction should, however, significantly help at the band edges since it is able to remove power here that the wide-band filter cannot. Alternatively, we might expect that a different choice of window function in the filtering step should give dramatic improvements at high/low redshifts when compared to the standard pipeline. \footnote{Alternatively, \citet{Liu2014FGB} describe a method for quantifying the leakage between k-bins from choice of window function therefore reducing emphasis on which window function is used; we do not explore this effect here.}

The PAPER power spectrum pipeline uses the Blackman-Harris window function when applying the iterative deconvolution filter, which gives the result of maximum foreground avoidance at band center. This choice of window function, however, produces limited improvements at the band edges due to down-weighting prior to delay transforming. Typically, if one were to forgo the use of a window function (i.e. using a Top-hat window) in the filtering of data with discontinuities this would result in significant spectral leakage, but this is not the case here. We can overcome spectral leakage in this instance because we are deconvolving both the window and sampling function from our filtered visibility. There are however limitations in doing this because deconvolution is a computationally demanding task. Furthermore, the product of the window and sampling functions can be complicated, and varies from visibility to visibility, so in some cases the iterative nature of this filtering may fail to converge. We therefore might expect the Top-hat window function to be an improvement over the Blackman-Harris near the band edges because it equally weights every frequency channel (as long as convergence in the iteration is reached). 
The points in Figure \ref{fig:threezdiff} are a difference of power spectra demonstrating the effect of window function choice in the wide-band filter (more positive is better).  For these points, the foreground filtering+subtraction analysis was applied in two cases (green; orange) to show the effect without a Top-hat window function used in the wide-band transform and after using the Top-hat window. We also show the change from Blackman-Harris to Top-hat (red) in a filtering only case to provide a point of reference for the inclusion of the foreground subtraction in the filtering+subtraction.
We see that the Top-hat filter leads to significant improvements at the band edges --- almost an order of magnitude larger near the horizon than the improvements we see from filtering+subtraction with a Blackman-Harris window. At the band center ($z = 8.4$), however, we see that the Top-hat filter results in a significantly worse result near the horizon: below $k$ of $\sim1.5  \ {\rm hMpc}^{-1}$, the Top-hat filter left significantly more residual foreground emission, making the difference negative. We expect that the Blackman-Harris should outperform the Top-hat window function in this case, since the Top-hat has more difficult sidelobes to deconvolve.

Lastly, the orange points in Figure \ref{fig:threezdiff} show the difference between filtering with a Blackman-Harris window and filtering+subtraction with a Top-hat window. In all three bands, we see that filtering+subtraction with a Top-hat leads to even more of an improvement than filtering with a Top-hat alone (red). At the band edges, the majority of the improvement over filtering with a Blackman-Harris appears to come from the switch to the Top-hat window, but the best results come when first subtracting a foreground model and then applying the wide-band deconvolution filter using a Top-hat window.  At the band center, the Blackman-Harris proves the best choice of window function in the wide-band filtering step. 

\subsection{Improved measurements within the horizon}
\label{sec:buffer}

\begin{figure}
\includegraphics[width=1.1\linewidth]{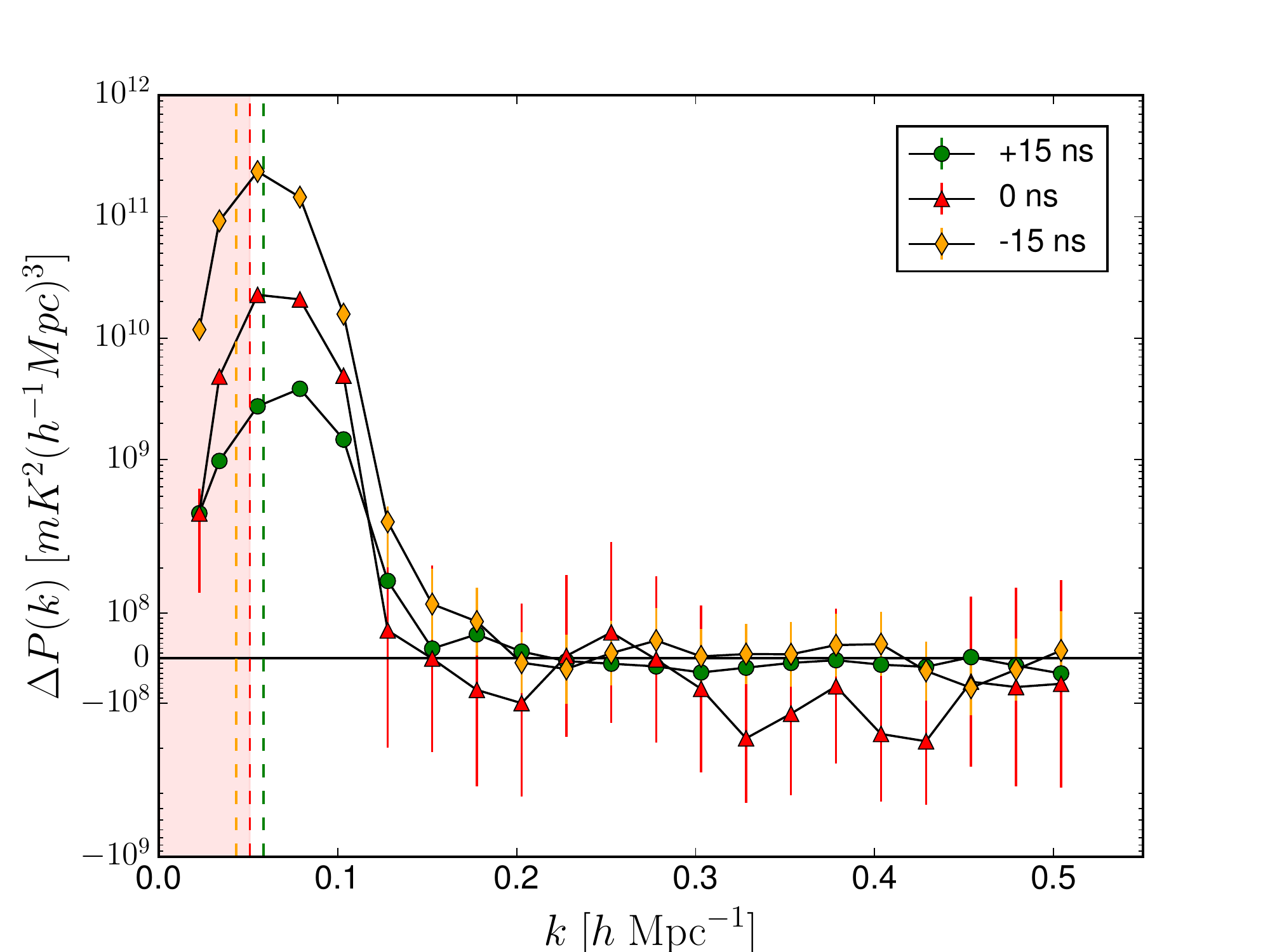}
\caption{Power spectra differences (filtering minus filtering+subtraction) for different sizes of filter: 15 ns beyond the horizon limit (green), at the horizon (red), and 15 ns inside the horizon (orange).  Smaller filters leave more of $k$ space available for EoR measurements (mode within the filter have effectively 100\% signal loss), and subtraction is seen to result in bigger improvements when smaller filter sizes are used.  With improved sky models, subtraction can therefore play an important role in maximizing the recovery of the 21 cm signal.}
\label{fig:NoBuffer}
\end{figure}

Delay filtering of foregrounds has been shown to be a powerful technique for 21cm data analysis, but it does have significant limitations. Modes within and directly outside the horizon (due to 15 ns extension) have effectively 100\% EoR signal loss when filtering and therefore become useless for cosmological measurements (see Appendix \ref{appendix:A}). This is of course unfortunate, as the EoR signal is brightest at these large scales. To begin to recover some of these modes, we can potentially use foreground subtraction prior to filtering and then reduce the filter's extent in $k$ back towards or even inside of the horizon. This means that we can maximize the effect of subtraction mitigating the foreground contamination by purposefully reducing the performance of filtering at the cost of sensitivity at higher $k$-modes. Figure \ref{fig:NoBuffer} shows power spectrum differences between filtering and filtering+subtraction for three choices of filter width: at the horizon (0 ns, red), 15 ns beyond the horizon, as used in the analysis above (green), and 15 ns inside of the horizon (orange).  We see that the improvements coming from the subtraction step grow larger the less aggressive the filter, i.e., filtering+subtraction results in the biggest improvement when the filter is set to 15 ns inside the horizon (-15 ns).  Overall, the total residual foreground power in $k$ modes near the horizon is minimized with the most aggressive filter (+15 ns), but as we continue to improve foreground models and the efficacy of subtraction, it will be worthwhile to revisit this analysis and see whether $k$ modes at or near the horizon can be turned into useful limits even in a foreground filtering approach. 

\vfill\null
\section{Hybrid Foreground Removal Technique Applied to MWA Phase I}
\label{section:mwa}

\begin{figure*}
	\includegraphics[width=1.0\linewidth,trim={0cm 0cm 0cm 0cm},clip]{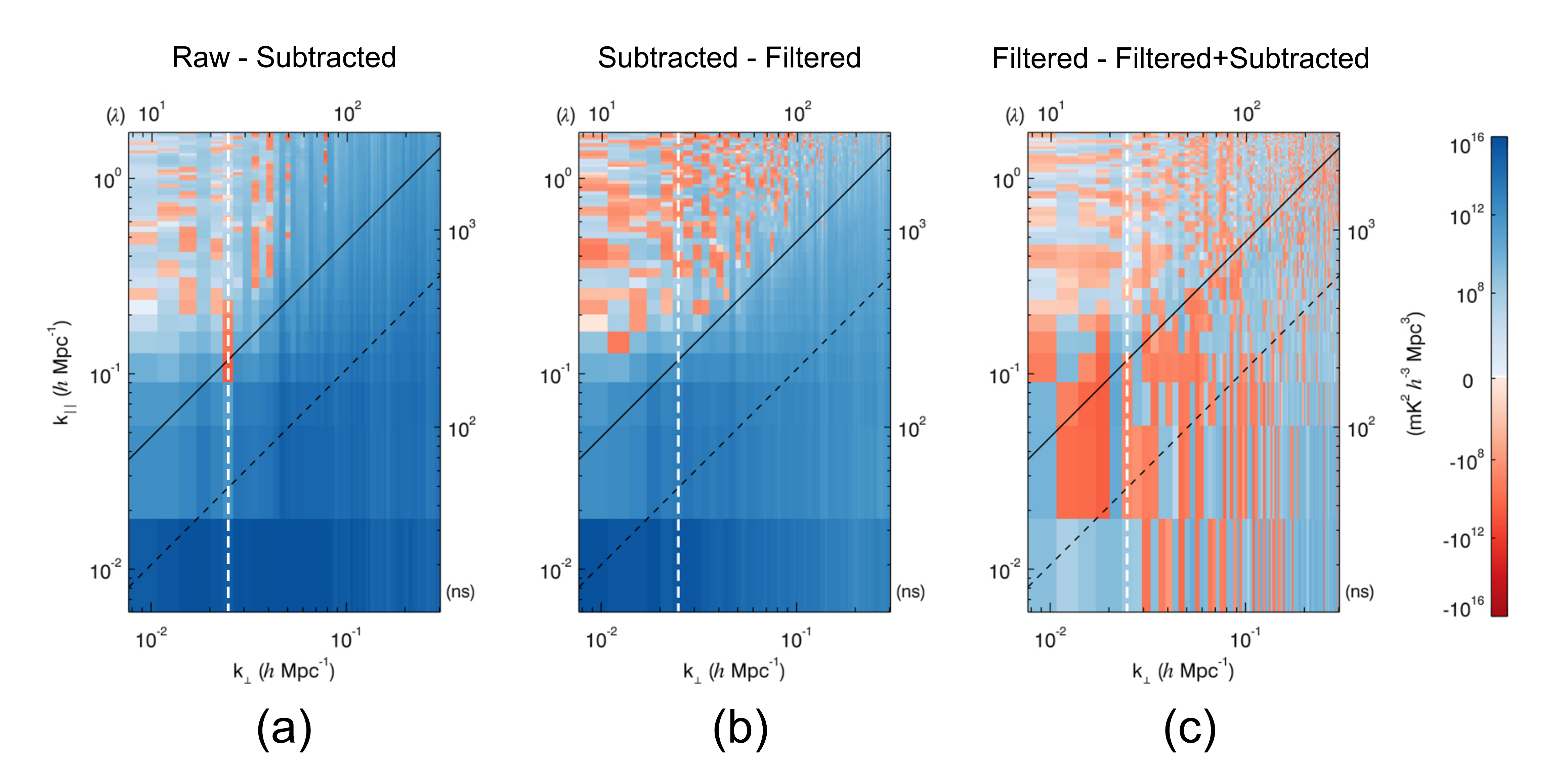}
    \caption{MWA 2D wedge power spectra difference plots showing the comparison between each foreground removal step where blue (red) indicates an improvement (worsening) between steps. The solid diagonal line represents the horizon limit while the broken line represents the Full Width Half Max of the MWA primary beam \citep{Sutinjo2015}. The vertical white broken line marks a 30 m baseline, equivalent in length to that used in the PAPER-64 analysis. The difference between the raw power spectrum with no foreground removal and one with FHD foreground subtraction is shown in (a) which gives an expected improvement inside of and beyond the horizon limit. The comparison between FHD foreground subtraction and wide-band filtering (b), shows an additional clear improvement over the standard MWA subtraction pipeline at the majority of $k_{\perp}$ modes nearest the horizon. Finally, comparing the use of filtering with both filtering+subtraction in panel (c), we do not see any clear improvement, but comparison with the PAPER-64 results suggest that we have reached a noise limit where any improvements are well below the sensitivity available with only three hours of data.}
\label{fig:DiffPSPECMWA}
\end{figure*}

To demonstrate the robustness and general applicability of the hybrid foreground removal technique across multiple instruments, we apply it to the three hours of MWA Phase I observations as described in Section \ref{sec:mwadata}. In many ways, these observations represent the extreme opposite of PAPER-64. The MWA Phase I has 128 tiles\footnote{MWA tiles consist of 16 individual dipole antenna elements that are steered with an analog beamformer, as opposed to the single dipole element used for drift scanning in PAPER-64.} laid out in a minimum redundancy configuration, giving it a much fuller $(u,v)$ coverage than PAPER-64. It also has baselines up to nearly 3 km in length, giving it a resolution of order a few arcminutes, a factor of $\sim 10$ improvement over PAPER-64.  FHD was also designed specifically for use with MWA interferometry data and the GLEAM catalog used for subtraction is a direct MWA product. We should thus expect better relative foreground subtraction when compared to PAPER-64. 

The results of applying the hybrid technique on approximately 3 hours of 
MWA Phase I observations at band center (182.415 MHz) are shown in 
Figure \ref{fig:DiffPSPECMWA} in `xx' instrument polarization. For the 
most accessible comparison to PAPER-64 hybrid foreground removal we 
continue to use the difference between power spectra but in the form of 
the cylindrically averaged 2D differenced power spectra. This gives us 
the ability to approximately compare similar MWA and PAPER 30m baseline 
difference power spectra and to view the effect of foreground 
filtering+subtraction across multiple MWA baseline types.
In these difference plots, blue indicates a region where the first step 
of the analysis has more power, and red indicates a region where the 
second step has more power.  In Figure \ref{fig:DiffPSPECMWA}(a) (left 
hand panel) we show the difference between raw data (calibrated, but no foreground subtraction) and the standard foreground subtraction used in the MWA power spectrum pipeline (as performed in e.g. \citealt{Pober2016}, \citealt{Jacobs2016}, and \citealt{Barry2016}).  We see, unsurprisingly, that foreground subtraction removes a significant amount of power inside the wedge, but there are also improvements well beyond the horizon limit (shown as a solid diagonal line). These improvements are on the same order of magnitude ($10^{16} \ \rm{mK}^{2}(h^{-1}\rm{Mpc})^{3}$ within the horizon) as seen in Figure \ref{fig:PowerSpectraMid} between the raw (green) and subtracted (red). Figure \ref{fig:DiffPSPECMWA}(b) compares foreground subtraction with the wide-band deconvolution foreground filtering.  Filtering removes more power within the horizon than subtraction, but we also see an improvement immediately outside of the horizon limit across all $k_{\perp}$. Note that no ``buffer" was used outside the horizon in the foreground filtering, so any power removed outside the horizon is almost certainly due to reduced spectral leakage.  At low $k_{\perp}$ and high $k_{\parallel}$ we approach a noise floor; since only three hours of data were used, it is considerably higher than the noise floor in the PAPER-64 analysis.
In Figure \ref{fig:DiffPSPECMWA}(c), comparing the filtered and filtered+subtracted power spectra, we see a similar noise level, but throughout all of $k$ space. If we assume our results for the MWA should closely mirror those from PAPER-64, then improvements at band center due to filtering+subtraction should be on the order of $\sim  10^{9}  \ \rm{mK}^{2}(h^{-1}\rm{Mpc})^{3}$ which is below the noise level we see here. 

\begin{figure}
	\includegraphics[width=1.05\linewidth]{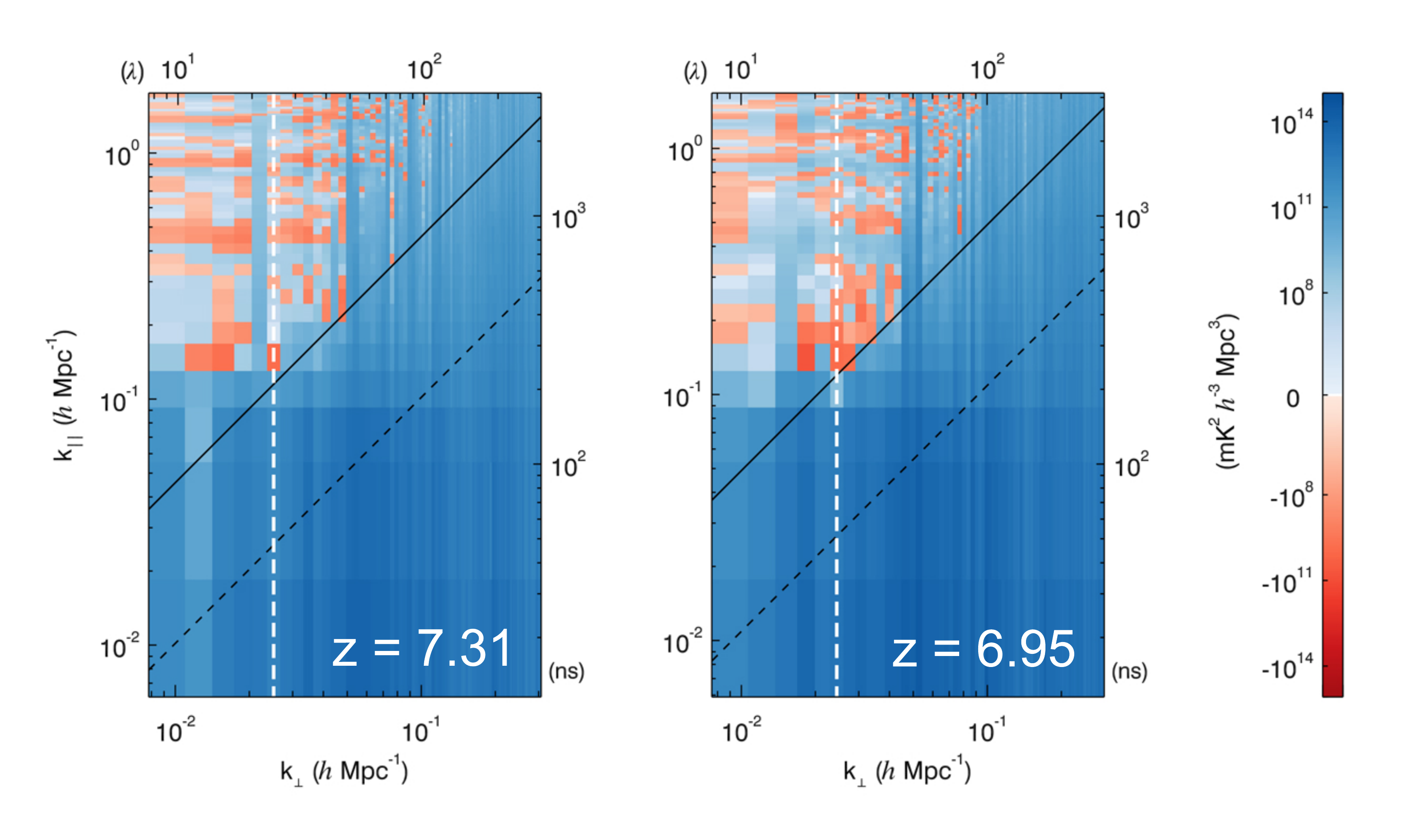}
    \caption{MWA 2D power spectra difference plots between filtering and filtering+subtraction at low band (7.31) and high band (6.95).  The results are generally similar to the equivalent PAPER-64 difference power spectra at the band edges. Improvements near the horizon are due to subtraction compensating for the down-weighting of band edges when windowing.}
    \label{fig:DiffPSPECMWAEdges}
\end{figure}

We also see similar results to the PAPER-64 analysis when looking at the MWA band edges (high/low redshift).  Based on the PAPER-64 analysis, we would expect improvements of $\sim 10^{12}-10^{13}  \ \rm{mK}^{2}(h^{-1}\rm{Mpc})^{3}$ from using filtering+subtraction over just filtering. This is consistent with the results shown in Figure \ref{fig:DiffPSPECMWAEdges}, which has a uniform decrease in power within the horizon and a slight, though noisy, improvement in the window.  While we have shown a Top-hat to be a much improved wide-band window for analysis near the band edges, MWA visibilities have a 2 s time resolution, compared to 10 s with PAPER.  This lower SNR coupled with a Top-hat window can make it more difficult for the iterative deconvolution filter to converge. Using a wide-band Blackman-Harris window function, even when analyzing data near the band edges, is therefore important to help convergence in this instance.

\section{Conclusions}
\label{section:conclusions}
The next generation of radio interferometers attempting to make a statistical detection of the 21cm EoR power spectrum will require the removal of as much foreground power as possible. To date, even the most promising foreground removal techniques leave behind residual contamination which is well above the most optimistically bright EoR signals. In this work, we have presented a hybrid technique which combines both model based foreground subtraction and delay based foreground filtering. In doing so, we have demonstrated the following:
\begin{itemize}
\item Even in parts of the EoR window, residuals from foreground filtering exist which are brighter than an EoR signal. 
\item Foreground subtraction can be successfully applied to arrays with sparse $(u,v)$ coverage (e.g. PAPER-64) using source catalogs derived from other arrays with high $(u,v)$ coverage (e.g. the MWA). The limiting factor in our ability to perform this subtraction is the numerical fidelity with which we can forward model visibilities for the large (1088 hour) PAPER data set.
\item Combining filtering and subtraction results in statistically significant improvements across the entire band, with the greatest improvements at band edges.
\item The choice of window function is an important ingredient in wide-band iterative deconvolution foreground filtering, especially when analyzing data at frequencies away from the band center.
\item Foreground subtraction can potentially enable less aggressive foreground filtering (i.e. removal of fewer modes near/inside the horizon), allowing for EoR measurements within the horizon.
\item Despite a more complicated frequency response (including the coarse band harmonics) and beam shape, foreground filtering can be successfully applied to MWA visibilities and results in a significant improvement over subtraction alone.
\item Improvements with filtering+subtraction are more readily seen as SNR improves, therefore we should expect for any improvements to sensitivity to become more visible with longer integration times.
\end{itemize}

The EoR signal is brightest at low $k$-modes, so foreground removal nearest (or inside) the horizon limit can have the greatest impact on overall sensitivity \citep{Pober2016}. We have seen that subtraction with a reduction in the net power prior to filtering can reduce leakage into the modes right outside the horizon. It would therefore be reasonable to subtract as much foreground power --- both diffuse and point sources --- prior to filtering for the most minimally contaminated EoR power spectrum estimates across all redshifts. We demonstrate this hybrid foreground subtraction and filtering technique as a proof of concept, and with additional computational resources to increase the $uv$-plane resolution to below the 0.35 $\lambda$ used here, we expect that this technique could be drastically improved upon. This would alleviate the need to low-pass filter our model visibilities and should result in near-horizon improvements by both modeling the spectral response of the instrument and reducing the overall dynamic range that the iterative deconvolution filter needs to overcome.

Furthermore, with an improved sky and instrument model we should expect subtraction to provide a more consistent improvement across all visibilities and not just ones containing our brightest foreground sources (Fornax A and Pictor A). More specifically, as weak and smaller sources are included in the analysis we approach a `subtraction floor' due to position errors between PAPER observations and the GLEAM catalog. This leads to improper or partial subtractions that result in no net removal of foreground power. Therefore improvements resulting from these changes in the case of PAPER-64, would stem directly from improved catalog positions and flux measurements, and the introduction of a diffuse model; but not necessarily from the additional modeling and subtraction of more weaker ($<$ 1 Jy) point sources.

Implementing the iterative deconvolution wide-band filter can be a computationally expensive process, especially when compounded with foreground subtraction which itself is laborious. The filtering step can also be accomplished without the use of the iterative deconvolution filter by doing a more traditional filtering in the delay domain. A sparse sampling function with missing data will however give significant ringing in the visibility. Ringing occurs when a signal is band-limited, which is approximately the case for a raw visibility where the foregrounds are $> 10^5$ when compared to anything beyond the horizon. This ringing is substantially reduced when subtraction is applied prior, potentially bringing the residual foregrounds to a point where they are no longer within the approximate band-limited regime. If subtraction can remove enough power to enable the use of an FIR foreground filter, the computational load and pipeline processing time can be significantly reduced. 

Future work will rely on applying this hybrid foreground removal approach to the newest season of observations from the Hydrogen Epoch of Reionization Array (HERA). Hardware improvements for HERA-37 alone should increase the sensitivity by a factor of $\sim$5 over PAPER-128 \citep{Deboer2017}. HERA's beam in terms of complexity positions it between PAPER and MWA, while having better horizon suppression than either of them \citep{Nithya2015}. This places additional importance on mitigating residual foreground contamination and puts us in a new regime for testing the hybrid foreground removal limitations as sensitivities increase. 
The inclusion of this hybrid approach to our most sensitive 21cm EoR interferometer should lead to the tightest, least contaminated power spectrum constraints over the redshift range of 7 $<$ z $<$ 11.

\section{Acknowledgments}
\acknowledgments
This material is based upon work supported by the National Aeronautics and Space 
Administration under the NASA Rhode Island Space Grant Consortium. 
JRK, JCP, and ARP would like to acknowledge the support from NSF Grants No. \#1440343 \& No. \#1636646. This material is additionally based upon work supported by the NSF under Grant No. \#1506024. 
APB is supported by an NSF Astronomy and Astrophysics Postdoctoral Fellowship under award \#1701440.
AL acknowledges support for this work by NASA through Hubble Fellowship grant \#HST-HF2- 51363.001-A awarded by the Space Telescope Science Institute, which is operated by the Association of Universities for Research in Astronomy, Inc., for NASA, under contract NAS5-26555.
This scientific work makes use of the Murchison Radio-astronomy Observatory, operated by 
CSIRO. We acknowledge the Wajarri Yamatji people as the traditional owners of the 
Observatory site. Support for the operation of the MWA is provided by the Australian 
Government (NCRIS), under a contract to Curtin University administered by Astronomy 
Australia Limited. We acknowledge the Pawsey Supercomputing Centre which is supported by 
the Western Australian and Australian Governments.
Additionally, this research was conducted using computational resources and services at 
the Center for Computation and Visualization, Brown University.

\bibliographystyle{aasjournal.bst}
\bibliography{references.bib}

\appendix
\section{Wide-band Deconvolution Filter Signal Loss}
\label{appendix:A}

\begin{figure*}[h]
\centering 
\subfigure[Blackman-Harris]{\label{fig:siglossbh}\includegraphics[width=90mm,trim={0cm 0cm 0cm 0cm},clip]{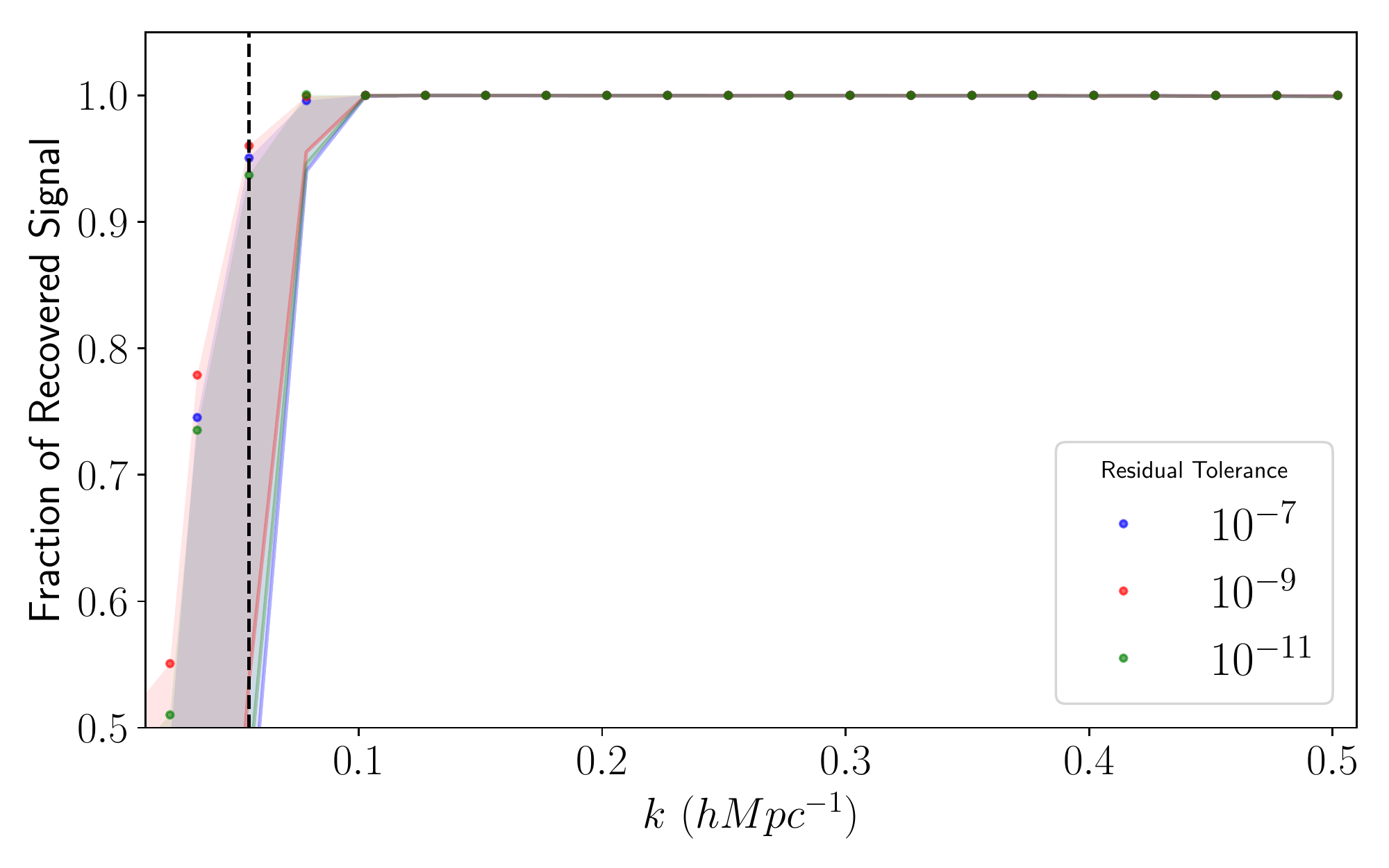}}%
~
\subfigure[Tophat]{\label{fig:siglossrect}\includegraphics[width=90mm,trim={0cm 0cm 0cm 0cm},clip]{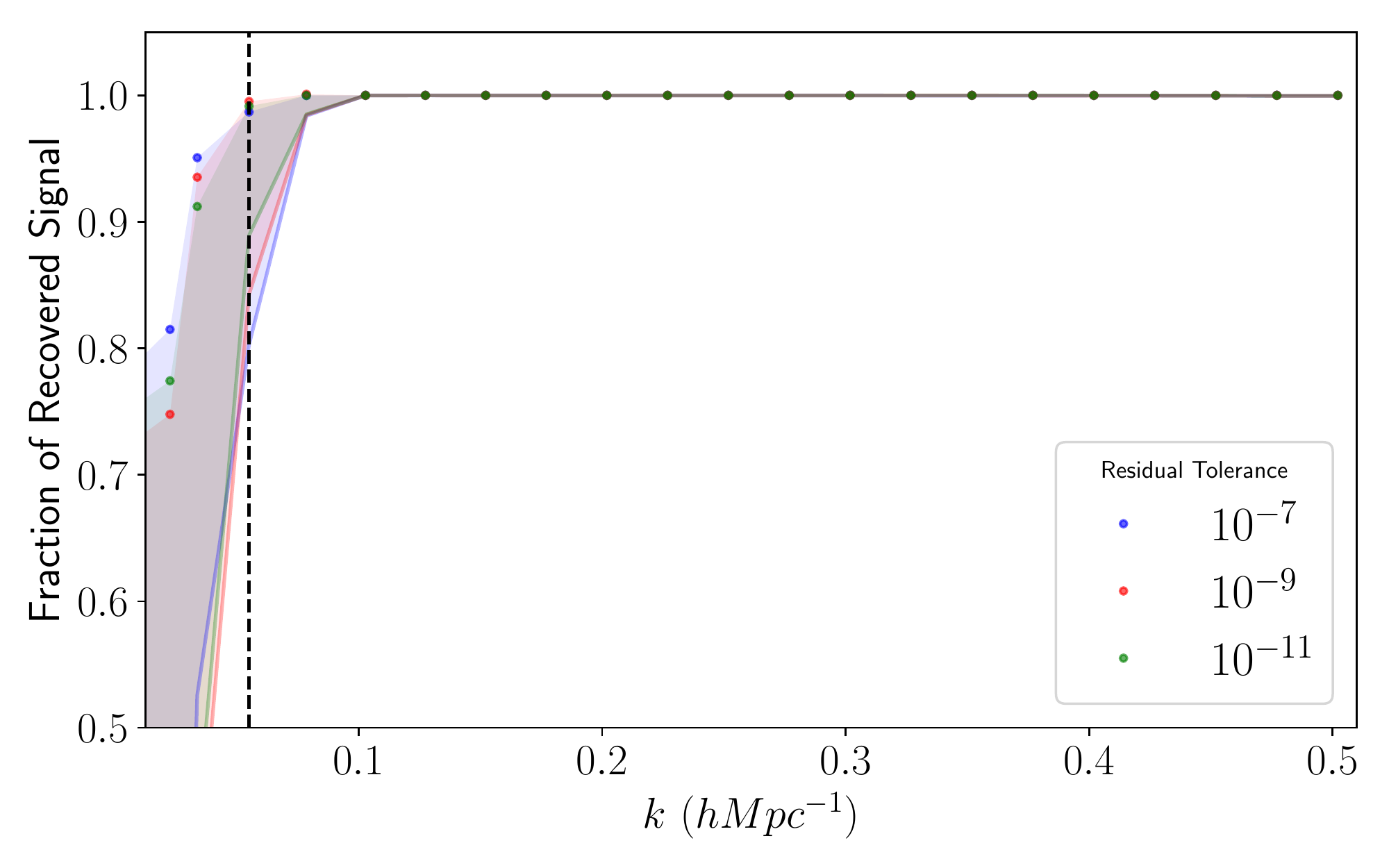}}

\subfigure[Across Baselines]{\label{fig:siglossbsl}\includegraphics[width=90mm,trim={0cm 0cm 0cm 0cm},clip]{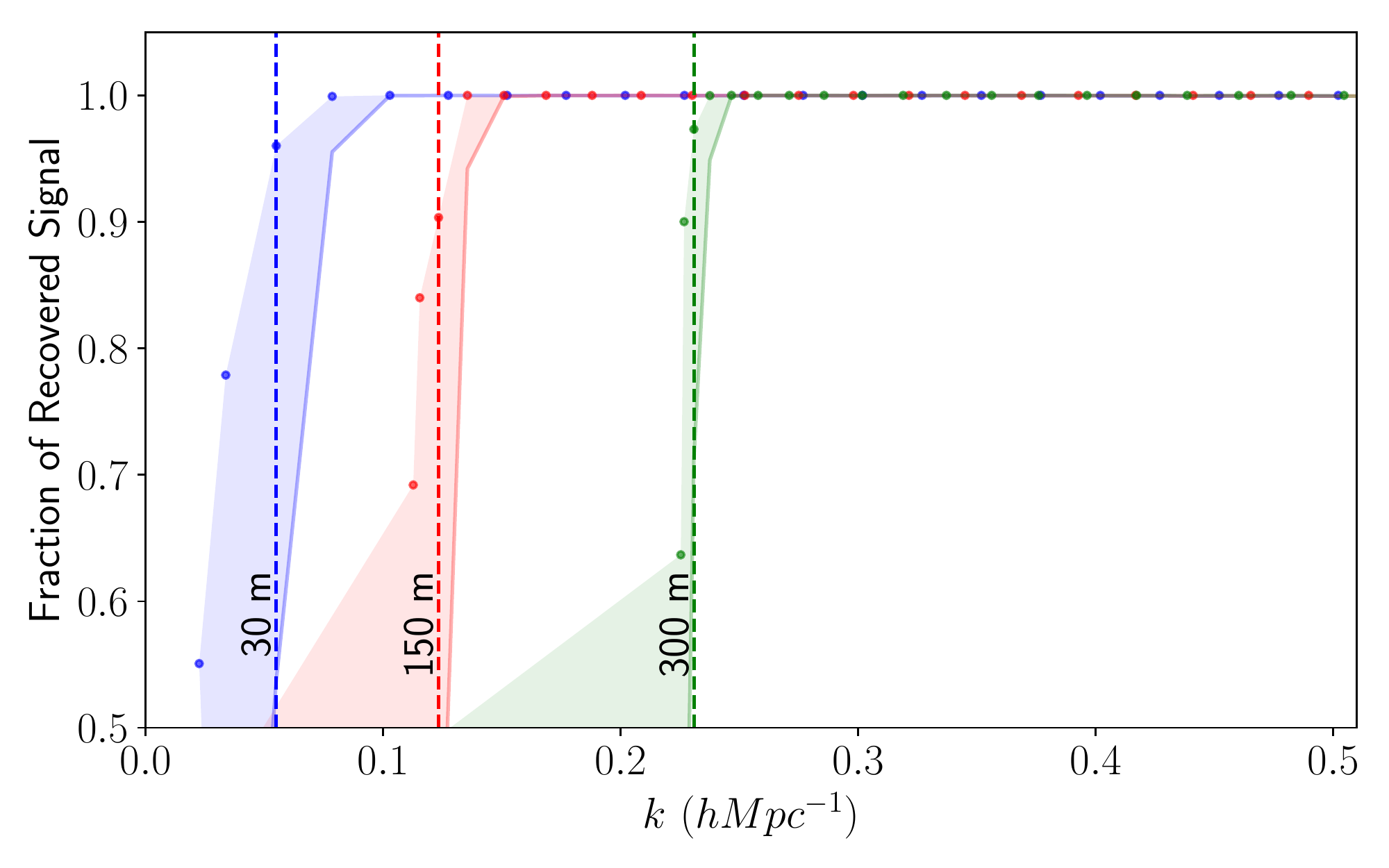}}

\caption{(a) The narrow-band power spectrum signal recovered after performing the wide-band iterative deconvolution filter using a Blackman-Harris window function. The dots represent the average signal recovered at each k-mode and the shaded region is the 3-$\sigma$ limit for our worst case signal recovery. Using the filtering residual tolerance of $10^{-9}$ we see that for the $k$-mode right outside the filter limit at $k = 0.079 \ hMpc^{-1}$ we can expect an average signal loss of $8.2\times 10^{-2}$\% with a 3-$\sigma$ upper-limit of $4.6$\%. (b) We approach the same analysis as (a) however we swap out the Blackman-Harris window with a Top-Hat window. The typical signal loss of the mode nearest the filter limit at $k = 0.079 \ hMpc^{-1}$ is $8.4\times 10^{-2}$\% with a 3-$\sigma$ upper-limit of $1.6$\%. (c) Although the PAPER-64 analysis consists of only one baseline type (30m) we also want to demonstrate the fraction of signal recovered over different length baselines (wider horizon filter limits) which is important to our MWA foreground subtraction and filtering analysis. The residual tolerance is $10^{-9}$ with a Blackman-Harris window applied, and the filtering is performed over baselines of the length 30m, 120m, and 300m. We show that the amount of signal recovered should essentially be a translation in $k$ when compared to (a) \& (b).}
\label{fig:SignalLoss}
\end{figure*}

The iterative wide-band deconvolution filter has the unfortunate side effect of destroying information at $k$-modes within our ``EoR Window'' nearest the horizon limit. The original analysis done in \citet{Ali2015} approached signal loss using a Blackman-Harris window function and a residual threshold (ratio of final to initial delay-space component) of $10^{-9}$. We extend the original analysis by quantifying the signal loss at all $k$-modes accessible by the PAPER-64 instrument for both the Blackman-Harris and Top-hat windows using several values of the residual tolerance. This is by no means an exhaustive analysis, but covers the range of filters used in this paper and other similar analyses from the PAPER collaboration. The amount of signal recovered is identified by injecting a known `EoR-like' signal ($V_{\rm{eor}}$) to a simulated PAPER-64 baseline visibility containing foregrounds ($V_{\rm{fgs}}$).
The visibility including the injected EoR signal $V_{\rm{fgs+eor}}$ and $V_{\rm{fgs}}$ are then wide-band filtered. We then create the power spectra $P_{\rm{fgs}}$ and $P_{\rm{fgs+eor}}$ and we compare the recovered signal in the form of

\begin{equation}
\bar{Y}(k)_{\rm{\small Frac. \ of \ Signal \ Rec.}} = \Bigg \langle\frac{\rm{Filtered}(P_{eor+data}) - \rm{Filtered}(P_{data})}{P_{eor}}\Bigg \rangle .
\label{eqn:signalloss}
\end{equation}
where $\langle \cdot \rangle$ represents the ensemble average and $\bar{Y}(k)_{\rm{\small Frac. \ of \ Signal \ Rec.}}$ is the ensemble average over the fraction of signal that is recovered from the wide-band filtering process.
We can get an idea of the typical (average) or extreme limit (3$\sigma$) for signal loss by looking at the ensemble average of recovered signal to injected signal over N = 1000 trials where we vary both injected signal realization and power ($10^{-10}\cdot P_{data} \leq P_{eor} \leq 10^{-4}\cdot P_{data}$). The fraction of signal recovered when using a Blackman-Harris window and Top-hat window is shown in Figure \ref{fig:siglossbh} and Figure \ref{fig:siglossrect}. The Top-hat window outperforms the Blackman-Harris with respect to recovering signal in the mode nearest the horizon filter limit ($k \approx 0.079 \ hMpc^{-1}$) because we sacrifice dynamic range for reducing correlation between adjacent $k$-modes. While both signal loss analyses in Figure \ref{fig:siglossbh} and Figure \ref{fig:siglossrect} demonstrate fairly similar worst case ($\bar{Y}$ + $3\sigma)$ signal loss across the tolerances demonstrated, its important to note that the magnitude of foregrounds removed from within the filter limit and the number of iterations to convergence (time) varies between tolerance. This leaves us with the optimal choice of $10^{-9}$ for our iterative wide-band deconvolution filter.

We also investigate how the increasing width of the filter with increasing baseline length affects signal loss.  The results of applying the same signal loss analysis but now to baselines of lengths 30m, 120m, and 300m are shown in Figure \ref{fig:siglossbsl}. We see that moving the filter limit gives us very similar results as the standard 30m filter limit, simply pushed further in $k$: modes inside the horizon limit have very large values of signal loss, while modes outside the horizon are essentially unaffected.  This analysis is useful in understanding the MWA 2D power spectra which contain different length baselines, seen in Figure \ref{fig:DiffPSPECMWA} and Figure \ref{fig:DiffPSPECMWAEdges}.



\section{Creating Model Visibilities with Fast Holographic Deconvolution}
\label{appendix:B}
\begin{figure}
	\includegraphics[width=1.\linewidth]{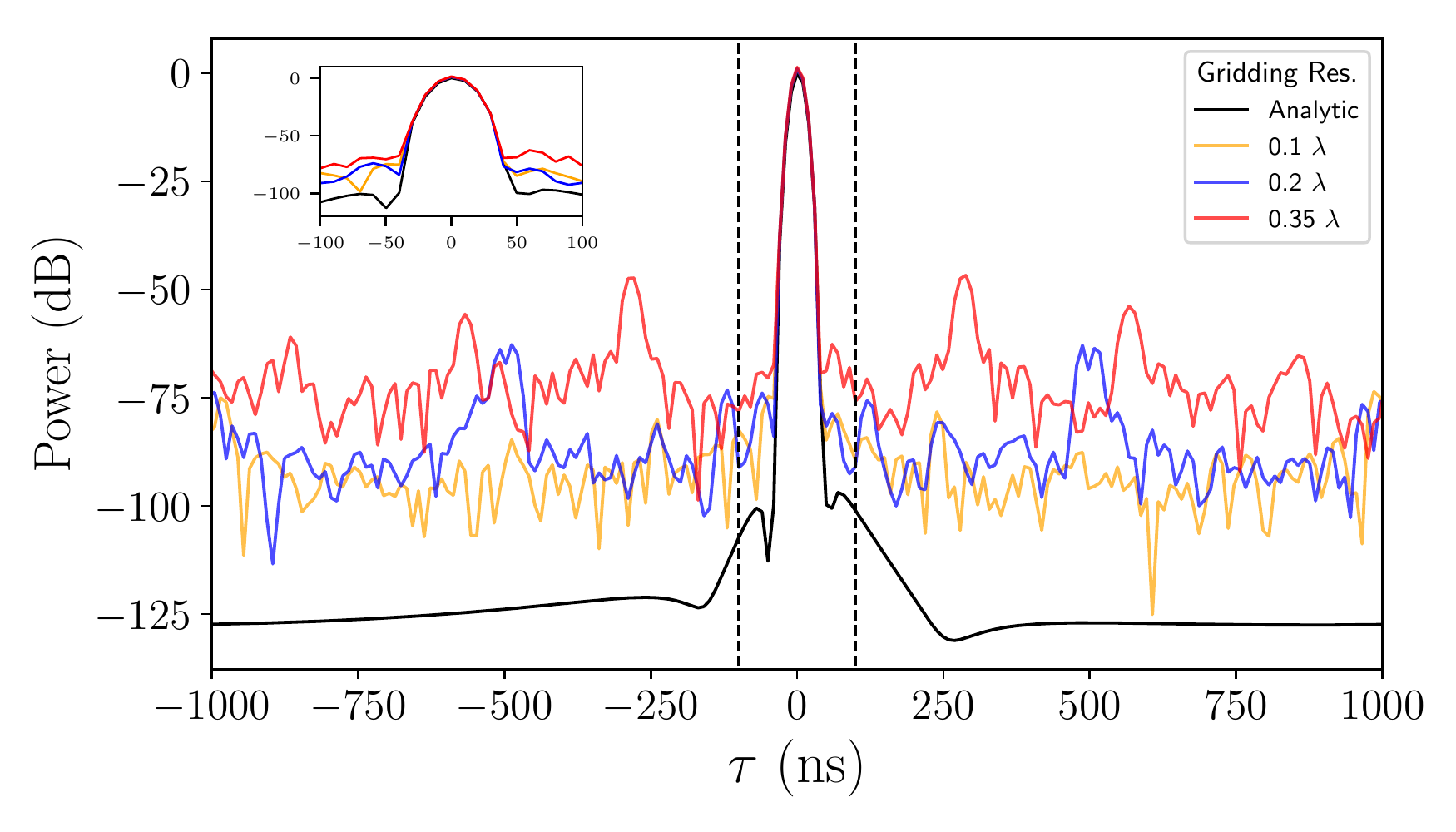}
    \caption{Simulation of a single point source in a 30 m baseline visibility at zenith for different $uv$-plane resolutions in FHD: 0.1 $\lambda$ (orange), 0.2 $\lambda$ (blue), and 0.35 $\lambda$ (red).  We also show the analytic response we expect for the same source (black). The inset figure shows the normalized peak spectra relative to the analytic source. At smaller $uv$-gridding resolutions, we see the power at high delays decreases while the peak source brightness converges to the analytic source. 
These model delay spectra examples also demonstrate a clear aliasing floor and thus lead to our motivation for filtering out high delays prior to subtraction.}
    \label{fig:single_source_fhd}
\end{figure}

\begin{figure}
	\includegraphics[width=1.\linewidth]{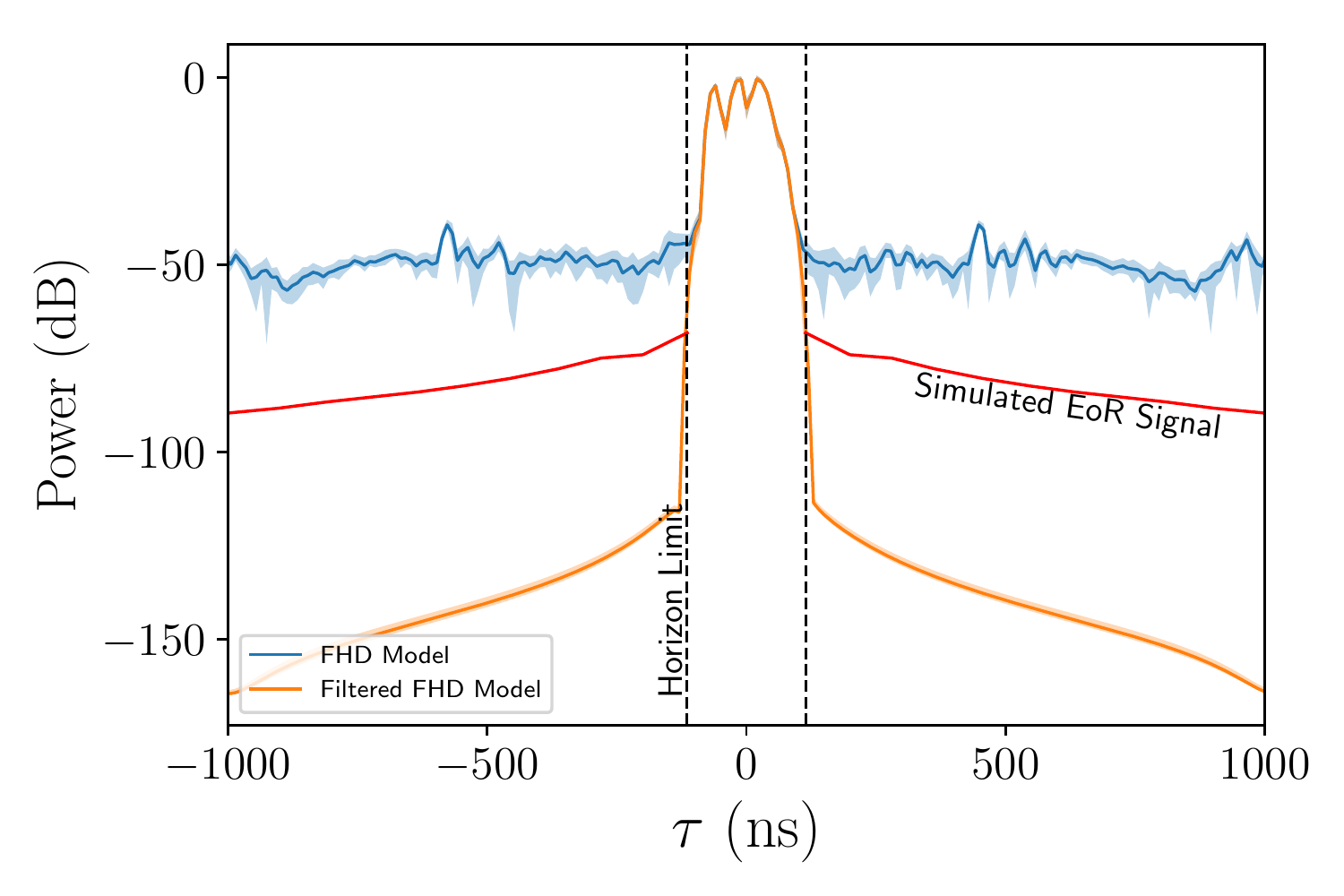}
    \caption{A many sources example (compare to Figure \ref{fig:single_source_fhd}) of a delay transformed and peak
power normalized FHD model visibility ($k_{res} = .35 \lambda$) for a 30 m baseline averaged over a single observation (blue), and the same model after
low pass filtering (orange) both with shaded regions demonstrating
the variance over time. A Blackman-Harris window
function has been applied in both cases. In the unaltered
curve, large amplitude spectral structure can be seen outside
of the delay horizon, $>$ 115 ns. As described in the text,
the origin of this high delay power is not physical. A fiducial
21cm Power Spectrum at 50\% reionization (Lidz et al. 2008)
(red) is shown as reference to demonstrate that the FHD
model must be filtered prior to subtraction so that our EoR
signal is not contaminated.}
    \label{fig:all_sky_sources}
\end{figure}

In this work, we use FHD \citep{Sullivan2012} to create model visibilities from a foreground source catalog and instrument model.  These model visibilities are then directly subtracted from the visibility data to perform our foreground subtraction step.  The FHD software package can perform many different operations, including calibration, image-based deconvolution, and visibility simulation; use of FHD as a visibility simulator has been presented in \cite{Barry2016} and \cite{Pober2016}.  Here, we describe the key steps in FHD visibility simulation to understand where the numerical artifacts described in Section \ref{sec:excess-power} appear.

When using a point source catalog as input, FHD uses a Discrete Fourier Transform (DFT) to take a source at a floating point location in (RA, Dec) to a discretized $uv$ plane.  Both the size of pixels in this $uv$ plane and the number of pixels (i.e. the maximum extent of the $uv$ plane) are free parameters in FHD.  All $uv$ planes for each point source being simulated are summed to create a model $uv$ sky-brightness distribution.  Model visibilities for each baseline are then created by integrating the product of the $uv$ response function (the Fourier transform of the primary beam from image domain to $uv$ space; \citealt{morales_and_matejek_2009}) at the $(u,v)$ coordinate of the baseline of interest and the model $uv$ sky-brightness distribution.  Because the $uv$ plane has been discretized into pixels, however, this integral is approximated as a discrete sum and can introduce numerical errors, particularly as a function of frequency (since the $(u,v)$ coordinates of a baseline are frequency dependent).

To illustrate these errors, we show the delay spectrum of a FHD-generated model visibility for a 30m East-West baseline and a single source in Figure \ref{fig:single_source_fhd}.  Different colored curves correspond to different pixel areas in the $uv$ plane in units of wavelength: 0.1 $\lambda$ (orange), 0.2 $\lambda$ (blue) and 0.35 $\lambda$ (red).  Increasing the $uv$-plane resolution clearly mitigates the high delay power created by spurious spectral structure, but comes at the cost of increased computational resources and would not be feasible without specialized high RAM machines. The black curve represents the analytic value assuming a Blackman-Harris window was used in the Fourier transform to delay space.  The gridding resolution 0.35 $\lambda$ is the resolution used in this analysis, and was used strictly for the purpose of efficiently calibrating, modeling, and subtracting our large dataset; running the full analysis at these higher resolutions is not computationally feasible, and even then, it does not bring the numerical error significantly below the level of the EoR signal (which is also approximately 80 dB below the foregrounds). Small flux density scaling errors between the different resolutions are a point of concern, but is largely mitigated in our analysis, as we effectively calibrate the flux scale of the FHD model to the absolutely calibrated PAPER data.

As noted in Section \ref{sec:excess-power}, this high-delay power is the limiting factor in our analysis.  If it is not removed, it sets a floor to best obtainable limits on the EoR signal with PAPER-64 data.  To remedy this problem we use our iterative deconvolution filter to low pass filter the 0.35 $\lambda$ model visibilities with filter horizon of 90 ns for our 30 m baselines.  Figure \ref{fig:all_sky_sources} illustrates the effect of this filter.  The delay spectrum of the FHD model prior to filtering is shown in blue and the delay spectrum after filtering is shown in orange.  For comparison, a model EoR signal is shown in red.  We acknowledge that this filter will also remove any \emph{real} high delay power (introduced by the instrument )\footnote{Because we model the sources on the sky as flat spectrum, they will not be a source of high delay power.} from the model visibilities.   Because the high delay power in the model is a limiting systematic, however, we choose to forgo the possibility of using the model to subtract high delay power from the data.  The fact that we still see improvements in the power spectrum (Section \ref{sec:bandcenter}) shows that we are reducing the overall power within the horizon to the point where our filtering process is not dynamic range limited.  Going forward, using either a modified version of FHD or another (likely more computationally intensive) code for simulating visibilities, we would not have to low pass filter our models and the improvement of the filtering+subtraction technique near the horizon should come from both effects: removing the chromatic footprint of the PAPER antenna response and reducing the dynamic range requirements on the delay filter.

\end{document}